\documentclass[pra,aps,superscriptaddress,twocolumn,floatfix,noshowpacs,nofootinbib,10pt]{revtex4-1}

\usepackage[utf8]{inputenc}
\usepackage{microtype}
\usepackage{amsmath,amssymb,mathrsfs}
\usepackage{graphicx}
\usepackage[colorlinks=true]{hyperref}

\DeclareMathOperator{\am}{am}
\DeclareMathOperator{\cn}{cn}
\DeclareMathOperator{\sn}{sn}

\begin{document}

\title{Supersolidity of cnoidal waves in an ultracold Bose gas}

\author{Giovanni I. Martone}
\affiliation{Universit\'{e} Paris-Saclay, CNRS, LPTMS, 91405 Orsay, France}
\affiliation{Laboratoire Kastler Brossel,
Sorbonne Universit\'{e}, CNRS, ENS-Universit\'{e} PSL,
Coll\`{e}ge de France; 4 Place Jussieu, 75005 Paris, France}

\author{Alessio Recati}
\affiliation{INO-CNR BEC Center and Dipartimento di Fisica,
Universit\`{a} di Trento, 38123 Povo, Italy}
\affiliation{Trento Institute for Fundamental Physics and Applications,
INFN, 38123, Trento, Italy}

\author{Nicolas Pavloff}
\affiliation{Universit\'{e} Paris-Saclay, CNRS, LPTMS, 91405 Orsay, France} 

\date{\today}

\begin{abstract}
  A one-dimensional Bose-Einstein condensate may experience nonlinear
  periodic modulations known as ``cnoidal waves''. We argue that such
  structures represent promising candidates for the study of
  supersolidity-related phenomena in a non-equilibrium state.
  A mean-field treatment makes it possible to rederive Leggett's formula
  for the superfluid fraction of the system and to estimate it analytically.
  We determine the excitation spectrum, for which we obtain analytical
  results in the two opposite limiting cases of (i) a linearly modulated
  background and (ii) a train of dark solitons. The presence of two Goldstone
  (gapless) modes -- associated with the spontaneous breaking of
  $\mathrm{U}(1)$ symmetry and of continuous translational invariance
  -- at large wavelength is verified. We also calculate the static structure
  factor and the compressibility of cnoidal waves, which show a divergent
  behavior at the edges of each Brillouin zone.
\end{abstract}

\maketitle

\section{Introduction}
\label{sec:introduction}
Supersolid phases of matter have attracted an increasing interest in
the last few years. In these configurations two apparently conflicting
properties, namely superfluidity and crystalline order, can coexist
giving rise to novel features~(see for instance the reviews~\cite{
Balibar_review,Boninsegni_review,Yukalov_review,Boettcher_review}).
The existence of such a phenomenon had initially been investigated,
and apparently ruled out, by Penrose and Onsager in the
1950s~\cite{Penrose1956}. It was reproposed shortly after by Gross.
In Refs.~\cite{Gross1957,Gross1958} he considered a system of interacting
bosons in the semiclassical limit, where the bosonic quantum field
can be replaced by a classical one, which obeys
a nonlinear field equation. The latter admits periodic solutions,
describing a uniform background with a crystal lattice on top of
it. In the subsequent decades the search for possible superfluid solid
phases was extended and other scenarios in which supersolidity could
occur were examined~\cite{Thouless1969,Andreev1969,Chester1970,
Leggett1970,Kirzhnits1971,Pitaevskii1984,Pomeau1994}. The main candidate
has been for many years the solid phase of helium. However, the most recent
experimental results and theory analyses seem to preclude superfluidity
in bulk solid helium~\cite{Balibar_review}, and the attention turned to
solid-helium two-dimensional films~\cite{Nyeki2017}.

On the other hand, significant progress has
been made with ultracold atomic gases starting from 2017, with the
first observations of an incompressible supersolid state in bosonic
systems coupled to two optical cavities~\cite{Leonard2017} and of
a superfluid smectic state in spin-orbit-coupled
Bose-Einstein condensates~\cite{Li2017}.  
Even more recently, coherent droplet arrays have been realized in
dipolar quantum gases~\cite{Tanzi2019a,Boettcher2019,Chomaz2019}.
This has stimulated a large amount of further experimental work,
shedding light on the spectrum and the
collective modes~\cite{Natale2019,Tanzi2019b,Guo2019,Petter2020}, the
superfluidity properties~\cite{Tanzi2019c}, and the out-of-equilibrium
dynamics~\cite{Ilzhoefer2019} of this exotic phase of matter.

A rather intriguing scenario for the occurrence of supersolidity is
the one pointed out by Pitaevskii in 1984~\cite{Pitaevskii1984}. He
proved that a sample of superfluid ${}^4$He flowing along a capillary
with a velocity exceeding Landau's critical value develops a layered
structure. This structure results from the condensation of excitations
close to the roton momentum~\cite{Iordanskii1980} and is at rest with
respect to the walls of the capillary. Its excitation spectrum is deformed
such that the system remains superfluid. These findings were later
confirmed by numerical simulations based on a density functional
approach~\cite{Ancillotto2005}. The same physics can be observed in
ultracold Bose gases as well, as found in recent times by Baym and
Pethick~\cite{Baym2012}. In this reference the authors assumed a
finite-range interaction between particles. This shifts the critical
momentum at which the Landau instability can occur to a finite
value. Similar to ${}^4$He, a large number of excitations with
momentum close to this value (called ``levons'') are created when
crossing the Landau velocity, which represents the onset of the
transition to the layered phase. The latter features a superfluid fraction
smaller than one.

In general, supersolid-like configurations can have smaller energy
than uniform ones only in special circumstances. Typically one needs
to have either specific kinds of interparticle interaction (such as
dipole-dipole, finite-range, or cavity-mediated), or a properly
modified single-particle spectrum (as in the case of
spin-orbit-coupled Bose-Einstein condensates). However, when none of
these conditions is fulfilled one may still have a supersolid behavior
in some excited state. This is the case of a standard
quasi--one-dimensional dilute Bose gas with repulsive contact
interaction, which is described by the Gross-Pitaevskii equation. This
equation is known to have spatially periodic stationary solutions,
which were studied by Tsuzuki in 1971~\cite{Tsuzuki1971}. Korteweg and
de Vries~\cite{Korteweg1895} had coined the term ``cnoidal waves'' for
solutions of this type, because they can be expressed in terms of the
Jacobi cosine amplitude function, denoted by
$\cn$~\cite{Abramowitz_Stegun_book}. In systems of bosons rotating in
a ring trap, transitions between metastable uniform and cnoidal
configurations have been
predicted~\cite{Kanamoto2008,Kanamoto2009}. Very recently,
cnoidal-wave--like solutions have been found for the extended
Gross-Pitaevskii equation describing a self-trapped cigar-shaped Bose
gas~\cite{Parit2020}.

Cnoidal waves can be regarded as the equivalent of Pitaevskii's
layered phase for an ultracold Bose gas. At variance with the case
considered in Ref.~\cite{Baym2012}, for repulsive contact interaction
the Landau instability takes place at vanishing momentum, and there is
no mechanism similar to levon condensation. Nevertheless, one can
achieve a cnoidal structure by moving an obstacle at a suitable
velocity through the condensate. The scope of this work is to
highlight that these configurations exhibit typical features of
supersolids in both their static and dynamic behavior. As such, they
are new candidates for studying phenomena related to supersolidity
within the most standard Bose-Einstein condensates. The latter do not
suffer from the strong three-body losses typical of dilute ultracold
systems in which the stabilization is due to beyond-mean-field equation
of states, as for dipolar gases and quantum mixtures. 

This article is organized as follows. In Sec.~\ref{sec:model} we
introduce the model to investigate and the equations governing it. In
Section~\ref{sec:cnoidal_sol} we present the derivation of the
cnoidal-wave solution and illustrate some of its most important
properties. The dynamic behavior of a cnoidal wave is discussed
in Sec.~\ref{sec:dyn_properties}, where we discuss the excitation
spectrum, the static structure factor and the compressibility.
We summarize in Sec.~\ref{sec:conclusion}. Finally, some technical
details are presented in the Appendices: Appendix~\ref{sec:cn_lim_cases}
presents the properties of the cnoidal wave in some limiting cases;
the procedure for solving the Bogoliubov equations is explained in
Appendix~\ref{sec:dyn_calculation}; and
Appendix~\ref{sec:dyn_lower_branch} computes the lower branch of the
spectrum of a train of dark solitons.

\section{The model}
\label{sec:model}
Let us consider a quasi--one-dimensional weakly interacting
Bose-Einstein condensate at zero temperature. The condensate wave
function $\psi(x,t)$ obeys the Gross-Pitaevskii equation
\begin{equation}
i \hbar \psi_t = - \frac{\hbar^2}{2m} \psi_{xx} 
+  (g |\psi|^2  - \mu) \psi \, .
\label{eq:GP}
\end{equation}
Here $m$ is the mass of a particle, $g > 0$ the interaction strength,
and $\mu$ the chemical potential. We use in Eq.~\eqref{eq:GP} and
throughout the paper the convention that subscripts denote derivatives
with respect to $x$ and $t$. 

Equation~\eqref{eq:GP} is a one-dimensional (1D) classical
field equation which is valid in the so-called ``1D mean-field
regime''~\cite{Menotti2002}. For a condensate transversely confined by
a harmonic trap of angular frequency $\omega_\perp$, this regime is
defined by the inequalities
\begin{equation}
\left(\frac{a}{a_\perp}\right)^2 \ll n_\mathrm{1D} \, a \ll 1 \, ,
\label{eq:GP_1Dmeanfield}
\end{equation}
where $a$ is the s-wave scattering length,
$a_\perp=\sqrt{\hbar/m \omega_\perp}$ is the transverse harmonic
oscillator length and $n_\mathrm{1D}$ is a typical order of magnitude of
the linear atom density $n=|\psi|^2$. In this regime one has
$g=2\hbar \omega_\perp a$~\cite{Olshanii1998}. For a transverse trap
of angular frequency $\omega_\perp=1$ kHz, one gets
$(a_\perp/a)^2=1.7\times 10^{-5}$ for $^{23}$Na, and
$(a_\perp/a)^2=2.6\times 10^{-4}$ for $^{87}$Rb, which means that the
1D mean-field regime where Eq.~\eqref{eq:GP} is valid ranges over
about 4 orders of magnitude in density.

We now perform a Madelung transform, which amounts to
writing the wave function under the form $\psi = A \, e^{i \Theta}$.
Inserting this expression into the Gross-Pitaevskii
equation~\eqref{eq:GP} yields two coupled equations for
the real quantities $A \geq 0$ and $\Theta$. The first one,
expressing the particle number conservation, is the continuity
equation, which reads
\begin{equation}
n_t + \left( \frac{\hbar \Theta_x}{m} \, n \right)_x = 0 \, ,
\label{eq:GP_td_phase}
\end{equation}
where we recall that $n = A^2$ is the linear density and 
$v = \hbar \Theta_x / m$
the velocity field. The second Madelung equation reads
\begin{equation}
\hbar \Theta_t = \frac{\hbar^2}{2m} \, \frac{A_{xx}}{A} - 
\frac{m}{2} \left(\frac{\hbar \Theta_x}{m}\right)^2 - g A^2 + \mu \, .
\label{eq:GP_td_amp}
\end{equation}
After taking the gradient on both sides, it becomes formally identical
to the Euler equation for the potential flow of an inviscid fluid,
with the addition of a ``quantum potential''.

\section{Cnoidal-wave solution}
\label{sec:cnoidal_sol}
The cnoidal-wave solution exhibited by the Gross-Pitaevskii
equation~\eqref{eq:GP} has been extensively studied in the
literature. In this section we review its derivation in order to fix
the notation and set the background for the subsequent calculations.
Then, we present some of its most relevant features, and in particular
we derive for the first time an analytic expression for the superfluid
density.

\subsection{Derivation of the cnoidal-wave solution}
\label{subsec:cn_sol_der}
\begin{figure}
\includegraphics[scale=1]{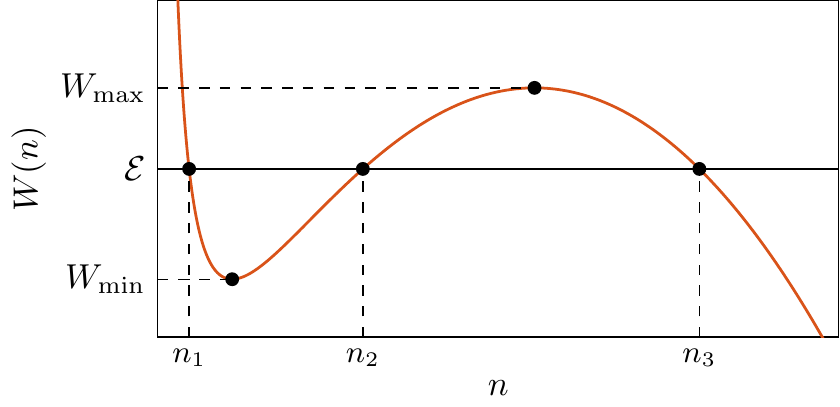}
\caption{Schematic behavior of the potential $W$ as a function of $n$.
The value of $W$ at the local minimum (maximum) is $W_\mathrm{min}$
($W_\mathrm{max}$). $n_1 \leq n_2 \leq n_3$ are the roots of
$W(n) = \mathcal{E}$.}
\label{fig:GP_pot}
\end{figure}

In order to find stationary solutions of the Gross-Pitaevskii
equation~\eqref{eq:GP} one has to set $n_t = 0$ and $\Theta_t = 0$.
This turns Eqs.~\eqref{eq:GP_td_phase} and~\eqref{eq:GP_td_amp} into
ordinary differential equations in $x$. Following Ref.~\cite{Tsuzuki1971}
we shall integrate these equations imposing the condition that the
condensate density and velocity oscillate in space at a given wavelength
$\Lambda$ around fixed average values $\bar{n}$ and $\bar{v}$.
Integrating once Eq.~\eqref{eq:GP_td_phase} with respect to $x$
one obtains
\begin{equation}
\Theta_x = \frac{m \mathcal{J}}{\hbar n} \, ,
\label{eq:GP_Sx}
\end{equation}
where $\mathcal{J}$ denotes the constant value of the current density.
We can use this result to eliminate $\Theta_x$ from
Eq.~\eqref{eq:GP_td_amp}. This yields
\begin{equation}
\frac{\hbar^2}{2m} A_x^2 +W(n) = \mathcal{E} \, ,
\label{eq:GP_Ax}
\end{equation}
where
\begin{equation}
W(n) = \frac{m \mathcal{J}^2}{2 n} - \frac{g n^2}{2} + \mu n \, .
\label{eq:GP_pot}
\end{equation}
Equation~\eqref{eq:GP_Ax} has the same mathematical structure as the
energy conservation of a classical particle having ``position'' $A$ at
``time'' $x$~\cite{Langer1967,Leboeuf2001}. The integration constant
$\mathcal{E}$ plays the role of the energy and $W$ that of the
external potential. 
In the following we assume that the current $\mathcal{J}$ verifies the
inequality~\cite{Mamaladze1966,Leboeuf2003}
\begin{equation}
\mathcal{J}^2 < \frac{8 \mu^3}{27 m g^2} \, ,
\label{eq:GP_pot_discr}
\end{equation}
which ensures that $W(n>0)$ has a local minimum $W_\mathrm{min}$ and a
local maximum $W_\mathrm{max}$, as illustrated in
Fig.~\ref{fig:GP_pot}. The maximal value~\eqref{eq:GP_pot_discr} of
$\mathcal{J}$ is analogous to the Ginzburg-Landau critical current in
a superconductor~\cite{deGennes_book}.

For given values of the parameters, the range of values that the
solutions of Eq.~\eqref{eq:GP_Ax} can take is fixed by the condition
$W(n) \leq \mathcal{E}$.  In particular, the roots of
$W(n) = \mathcal{E}$ identify the extrema of $n$, where $A_x = 0$.
They are the analogous of the turning points of a classical system. In
the case $W_\mathrm{min} < \mathcal{E} < W_\mathrm{max}$ 
considered in Fig.~\ref{fig:GP_pot} there are 3 such roots
which we denote as $n_1$, $n_2$ and $n_3$ with
$0 \leq n_1 \leq n_2 \leq n_3$.  Let us then rewrite Eq.~\eqref{eq:GP_Ax} as
\begin{equation}
\begin{split}
\left(\frac{n_x}{2}\right)^2
&{} = \frac{m g}{\hbar^2} \left( n^3 - \frac{2\mu}{g} n^2 
+ \frac{2 \mathcal{E}}{g} n - \frac{m \mathcal{J}^2}{g} \right) \\
&{} = \frac{m g}{\hbar^2} \left(n - n_1\right) 
\left(n - n_2\right)  \left(n - n_3\right) \, .
\end{split}
\label{eq:cn_dens_eq}
\end{equation}
Comparing the two rows of the above equation one immediately finds out
how to express $\mu$, $\mathcal{E}$, and $\mathcal{J}$ in terms of
$n_1$, $n_2$, and $n_3$. The result reads
\begin{subequations}
\label{eq:cn_ns}
\begin{align}
\mu &{} = \frac{g(n_1 + n_2 + n_3)}{2} \, ,
\label{eq:cn_ns_mu} \\
\mathcal{E} &{} = \frac{g(n_1 n_2 + n_2 n_3 + n_3 n_1)}{2} \, ,
\label{eq:cn_ns_E} \\
\mathcal{J}^2 &{} = \frac{g n_1 n_2 n_3}{m} \, .
\label{eq:cn_ns_J}
\end{align}
\end{subequations}
A bounded solution of Eq.~\eqref{eq:cn_dens_eq} oscillates between
$n_1$ and $n_2$ and thus is of the form
\begin{equation}
n(x) = n_1 \cos^2 \varphi(x) + n_2 \sin^2 \varphi(x) \, .
\label{eq:cn_dens_ansatz}
\end{equation}
Inserting this Ansatz into Eq.~\eqref{eq:cn_dens_eq} yields (upon
properly defining the spatial origin)
\begin{equation}
\varphi(x) = \am 
\left(\frac{\sqrt{m g (n_3 - n_1)}}{\hbar}\, x \middle| m_e \right) \, ,
\label{eq:cn_phi_sol}
\end{equation}
where $\am$ is Jacobi's amplitude function~\cite{Abramowitz_Stegun_book}
and
\begin{equation}
m_e = \frac{n_2 - n_1}{n_3 - n_1} \in [0,1] \, .
\label{eq:cn_param}
\end{equation}
The corresponding density and phase read
\begin{align}
n(x) &{} = n_1 + (n_2 - n_1) 
\sn^2 \left(\frac{\sqrt{m g (n_3 - n_1)}}{\hbar} \, x \middle| m_e \right) 
\, ,
\label{eq:cn_dens_sol} \\
\Theta(x) &{} = \pm \sqrt{\frac{n_2 n_3}{n_1 (n_3-n_1)}} 
\, \Pi\left(-n_e; \varphi(x) | m_e\right) \, .
\label{eq:cn_phase_sol}
\end{align}
Here $\sn(u|m_e)=\sin(\am(u|m_e))$ is the Jacobi sine amplitude
function, and $\Pi\left(-n_e; \varphi | m_e\right)$ denotes the incomplete
elliptic integral of the third kind~\cite{Abramowitz_Stegun_book}.
The quantity ${}- n_e = - (n_2-n_1)/n_1$ is called the ``characteristic''.
The condensate phase~\eqref{eq:cn_phase_sol} was determined integrating
Eq.~\eqref{eq:GP_Sx} with respect to $x$, imposing $\Theta(0) = 0$
for simplicity; the plus (minus) sign corresponds to a positive (negative)
value of the current $\mathcal{J}$.

Equations~\eqref{eq:cn_dens_sol} and~\eqref{eq:cn_phase_sol} express
the cnoidal-wave solution of the Gross-Pitaevskii
equation~\eqref{eq:GP}. It was first investigated by Tsuzuki in
Ref.~\cite{Tsuzuki1971}, see also Ref.~\cite{Carr2000}. This solution
depends on the three parameters $n_1 \leq n_2 \leq n_3$. It represents
a stationary layered structure, i.e., such that its density profile exhibits
periodic spatial modulations; a fixed current $\mathcal{J}$ flows through
the fringes. The oscillation wavelength and average density are computed
in Sec.~\ref{subsec:cn_sol_prop} below and are given by
Eqs.~\eqref{eq:cn_wavelength} and~\eqref{eq:cn_avg_dens}, respectively.
The modulations correspond to a spontaneous breaking of continuous translational
invariance. Because of the simultaneous presence of superfluid and crystal
order, cnoidal waves are expected to exhibit a supersolid behavior in both
their static and dynamic properties. These aspects will be elucidated in the
following sections.

\subsection{Properties of the cnoidal-wave solution}
\label{subsec:cn_sol_prop}
We shall now examine some characteristic features of cnoidal waves. These
include the average density, the contrast of the density modulations,
the superfluid fraction, and the energy per particle.

\subsubsection{Density profile and contrast of the fringes}
\label{subsubsec:cn_dens_prof}
The density profile~\eqref{eq:cn_dens_sol} oscillates with a wavelength
\begin{equation}
\Lambda = \frac{2 K(m_e) \hbar}{\sqrt{m g (n_3 - n_1)}} \, ,
\label{eq:cn_wavelength}
\end{equation}
where $K(m_e)$ is the complete elliptic
integral of the first kind~\cite{Abramowitz_Stegun_book}. These oscillations
occur around an average value given by~\cite{Tsuzuki1971}
\begin{equation}
\bar{n} = \frac{1}{\Lambda} 
\int_{-\frac{\Lambda}{2}}^{\frac{\Lambda}{2}} \! d x \, n(x) 
= n_1 + (n_3 - n_1) \left[ 1 - \frac{E(m_e)}{K(m_e)} \right] \, ,
\label{eq:cn_avg_dens}
\end{equation}
where
$E(m_e)$ is
the complete elliptic integral of the second
kind~\cite{Abramowitz_Stegun_book}. Using this average density we can
define the healing length $\xi = \hbar / \sqrt{m g \bar{n}}$ and the
sound velocity $c = \sqrt{g \bar{n} / m}$.
It is useful to rewrite $n_1$, $n_2$, and $n_3$ in terms of $\bar{n}$
and of the two dimensionless parameters $m_e$ and
\begin{equation}
\eta = \frac{n_3 - n_1}{\bar{n}} \, .
\label{eq:cn_eta}
\end{equation} 
From Eqs.~\eqref{eq:cn_param} and~\eqref{eq:cn_avg_dens} one gets
\begin{subequations}
\label{eq:cn_roots}
\begin{align}
\frac{n_1}{\bar{n}} &{} = 1 - \eta \left[1 - \Gamma(m_e)\right] \, , 
\label{eq:cn_roots_1} \\
\frac{n_2}{\bar{n}} &{} =1 + \eta \left[\Gamma(m_e) + m_e - 1\right] \, , 
\label{eq:cn_roots_2} \\
\frac{n_3}{\bar{n}} &{} = 1 + \eta \Gamma(m_e) \, ,
\label{eq:cn_roots_3}
\end{align}
\end{subequations}
where $\Gamma(m_e) = E(m_e) / K(m_e)$.
One can easily check that the conditions $0 \leq m_e \leq 1$ and
$\eta \geq 0$ are sufficient to ensure that $n_2$ and $n_3$ are
non-negative. Additional constraints come from the requirement
$n_1 \geq 0$. The latter is satisfied for any $0 \leq m_e \leq 1$ if
$0 \leq \eta \leq 1$; but, if $\eta > 1$, $m_e$ should not be larger
than a threshold value $m_e^\mathrm{max}$ defined by
$\Gamma(m_e^\mathrm{max})=(\eta - 1) / \eta$. In the following
we shall see that, when considered as functions of $m_e$,  the
various observables have different behaviors, depending on
whether $\eta$ is smaller or larger than $1$.

Making use of the average density Eq.~\eqref{eq:cn_avg_dens}
we can decompose the density~\eqref{eq:cn_dens_sol} into a
uniform and a modulated component as
$n(x) = \bar{n} + \Delta n(x)$, with
\begin{equation}
\Delta n(x) = \bar{n} \eta
\left[ m_e \sn^2 \left( \sqrt{\eta} \, \frac{x}{\xi} \middle| m_e \right)
+ \Gamma(m_e) - 1 \right] \, .
\label{eq:cn_mod_dens}
\end{equation}

\begin{figure}[htb]
\includegraphics[scale=1]{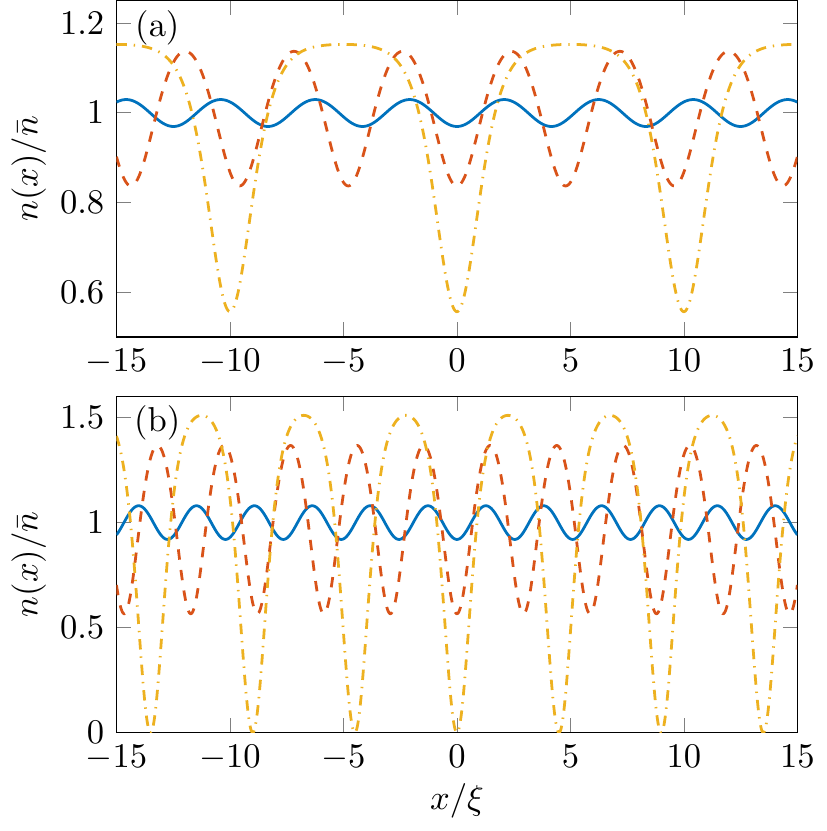}
\caption{Density profiles $n(x)$ of cnoidal waves
for (a) $\eta = 0.6$  and (b) $\eta = 1.6$ and for
$m_e = 0.1$ (blue-solid lines) and $m_e = 0.5$ (red-dashed lines).
The value of $m_e$ for the two yellow dash-dotted curves was chosen
so as to minimize the superfluid density~\eqref{eq:cn_sf_frac}
[see also the red dashed lines of
Figs.~\ref{fig:cn_cont_sf_frac}(c)-(d)]. It is equal to $0.993$ in
panel (a) and to $m_e^\mathrm{max}=0.943$ in (b).}
\label{fig:cn_dens_prof}
\end{figure}

In Fig.~\ref{fig:cn_dens_prof} we report a few density profiles
of cnoidal waves for different values of $m_e$ and $\eta$. 
At small $m_e$ the oscillations have small amplitude and are
practically sinusoidal, as discussed in Appendix~\ref{subsec:cn_lin}.
Increasing $m_e$ at fixed $\eta$ produces
fringes with larger amplitude and wavelength, as well as significant
deviations from the sinusoidal behavior. When $m_e$ is close to $1$
the density profile takes the characteristic shape of a ``soliton
train'', made by quasi-uniform regions separated by thin deep valleys.

\begin{figure}[htb]
\includegraphics[scale=1]{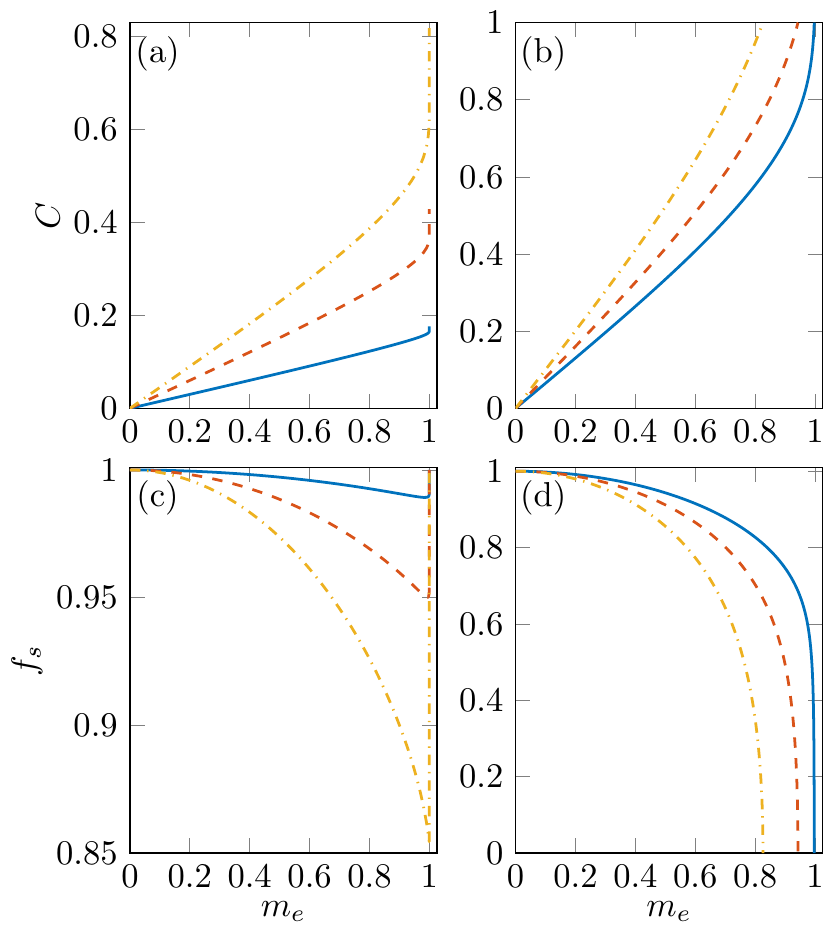}
\caption{Contrast of the fringes [(a)-(b)] and corresponding
superfluid density [(c)-(d)] as functions of $m_e$. In panels (a)
and (c) we take $\eta = 0.3$ (blue solid line), $0.6$ (red dashed
line), and $0.9$ (yellow dash-dotted line). Instead, in panels (b)
and (d) we choose $\eta = 1.3$ (blue solid line), $1.6$ (red dashed
line), and $2.0$ (yellow dash-dotted line). The corresponding values
of $m_e^\mathrm{max}$ are $0.997$, $0.943$, and $0.826$, respectively.}
\label{fig:cn_cont_sf_frac}
\end{figure}

A useful quantity to characterize the fringes is their contrast,
\begin{equation}
C = \frac{n_2 - n_1}{n_2 + n_1} =
\frac{m_e}{m_e+2\Gamma(m_e)-2(\eta-1)/\eta} \, .
\label{eq:cn_contrast}
\end{equation}
At small $m_e$ the contrast behaves like $C \simeq m_e \eta / 2$,
whereas beyond this regime two cases should be distinguished. When
$\eta \leq 1$ the parameter $m_e$ can vary between $0$ and $1$, the
two extreme values corresponding to a uniform and a dark-soliton
configuration, respectively (see Appendix~\ref{sec:cn_lim_cases}).
Consequently the contrast smoothly increases from $0$ to a value
$\eta/(2-\eta) \leq 1$ at increasing $m_e$ [see
Fig.~\ref{fig:cn_cont_sf_frac}(a)]. In particular for $m_e = \eta = 1$,
which corresponds to a black soliton, one has $C = 1$.
The situation is different for $\eta > 1$, where $m_e$ can only vary
in a smaller range of values, as discussed earlier. As shown in
Fig.~\ref{fig:cn_cont_sf_frac}(b), in this case the contrast always
reaches its maximum value $C = 1$ at $m_e=m_e^\mathrm{max}$, and thus
one can have density fringes oscillating between $0$ and
$\eta m_e^\mathrm{max} \bar{n}$ [see the yellow dash-dotted curve of
Fig.~\ref{fig:cn_dens_prof}(b)]. 

\subsubsection{Average velocity and superfluid fraction}
\label{subsubsec:cn_sf_frac}
The velocity field $v = \hbar \Theta_x / m$ oscillates with the same
wavelength $\Lambda$ as the density. Its average value is
\begin{equation}
\bar{v} = \frac{1}{\Lambda} 
\int_{-\frac{\Lambda}{2}}^{\frac{\Lambda}{2}} d x \, v(x)
= \frac{\mathcal{J}}{f_s \bar{n}} \, ,
\label{eq:cn_avg_vel}
\end{equation}
where we define 
\begin{equation}
\begin{split}
f_s & = \left[\frac{\bar{n}}{\Lambda} 
\int_{-\frac{\Lambda}{2}}^{\frac{\Lambda}{2}}
\frac{ d x}{n(x)} \right]^{-1} \\
&{} =
\frac{1 - \eta \left[1 - \Gamma(m_e)\right]}{\Pi(- n_e | m_e)}
 \,  K(m_e) \, ,
\end{split}
\label{eq:cn_sf_frac}
\end{equation}
and $\Pi\left(-n_e | m_e\right) = \Pi\left(-n_e; \pi/2 | m_e\right)$
is the complete elliptic integral of the third
kind~\cite{Abramowitz_Stegun_book}.  Notice that
Eq.~\eqref{eq:cn_avg_vel} can be rewritten in the natural form 
\begin{equation}
\mathcal{J} = f_s \bar{n} \bar{v} \, ,
\label{eq:cn_sf_current}
\end{equation} 
which indicates that $f_s$ is precisely the superfluid fraction of the
system~\cite{Svistunov_book}. Our first equality in Eq.~\eqref{eq:cn_sf_frac}
coincides indeed with the well-known estimate of the superfluid fraction
for a supersolid introduced by Leggett~\cite{Leggett1970,Leggett1998}.
Actually in these works the first row of
Eq.~\eqref{eq:cn_sf_frac} was shown to be an upper bound to the real
superfluid fraction. It was derived using an Ansatz wave function that
assumes all the particles in the superfluid to have the same
phase. This assumption is weaker than the one we make in the present
work using the Gross-Pitaevskii theory, in which all the atoms have
the same wave function. This is why within this approximation
Eq.~\eqref{eq:cn_sf_frac} is found as an exact result.

Equations~\eqref{eq:cn_avg_vel}, \eqref{eq:cn_sf_frac}, and
\eqref{eq:cn_sf_current}
constitute one of the important results of the present work,
where Leggett's formula for $f_s$ comes out as an immediate consequence
of the definition of the average condensate velocity, allowing us to provide
also an analytical expression for $f_s$.

It is worth investigating the behavior of $f_s$ as a function of the
parameters characterizing the cnoidal-wave solution. The strength of $f_s$
as a function of $m_e$ and for different values of $\eta$ is reported
in Figs.~\ref{fig:cn_cont_sf_frac}(c)-(d). At fixed $\eta$
and small $m_e$, where cnoidal waves reduce to Bogoliubov oscillations
(see Appendix~\ref{subsec:cn_lin}), the superfluid fraction retains the
trivial value $f_s = 1$. Expanding Eq.~\eqref{eq:cn_sf_frac} up to
second order in $m_e$, so to take the first nonlinear correction into
account, one finds $f_s \simeq 1 - \eta^2 m_e^2 / 8 = 1 - C^2/2$. This
result matches very well the curves in
Fig.~\ref{fig:cn_cont_sf_frac}(c)-(d) at small $m_e$ (an analogous
relation was recently derived in Ref.~\cite{Chomaz2020} for a shallow
sine-modulated supersolid). For $\eta < 1$
[Fig.~\ref{fig:cn_cont_sf_frac}(c)] $f_s$ decreases at increasing $m_e$
down to a minimum, that is typically attained at some $m_e$ very close
to $1$; then, it undergoes a smooth but very steep ascent and goes back
to $1$ at $m_e=1$, where the cnoidal wave turns into a dark soliton (see
Appendix~\ref{subsec:cn_sol}). Instead, when $\eta = 1$ the superfluid
density continues to drop down to $0$ as $m_e$ approaches $1$.
Also in the $\eta > 1$ regime [Fig.~\ref{fig:cn_cont_sf_frac}(d)] $f_s$
monotonously decreases with $m_e$ from $1$ to $0$, the latter
value being attained at $m_e=m_e^\mathrm{max}$, where the contrast of
the fringes~\eqref{eq:cn_contrast} is $1$. Thus a cnoidal wave with
strong modulations is very weakly superfluid, again in agreement with
Leggett's arguments~\cite{Leggett1970,Leggett1998}. 

On the theory side, the situation encountered here is common also
to the modulated configurations studied for dipolar Bose gases. Leggett's 
equation coincides with the superfluid density obtained from single-orbital
density functional theory, a.k.a. extended Gross-Pitaevskii equation,
and it becomes zero when the periodic structure has contrast $C=1$
(see, e.g., Refs.~\cite{Roccuzzo2019,Zhang2019,Chomaz2020}). Although a number
of properties have been experimentally measured, the smallness of the sample
and its short lifetime have precluded direct access to the superfluid density
so far (see however Ref.~\cite{Tanzi2019c} for a first try in this direction).  

Let us also mention that in the stripe phase of spin-orbit-coupled Bose gases
the maximum achievable value of the contrast depends on the interaction
strength in the various spin channels, and the deeply modulated regime with
a small superfluid fraction is more challenging to reach~\cite{Martone2014}.

\subsubsection{Energy per particle}
\label{subsubsec:cn_energy_part}
The energy per particle is given by
\begin{equation}
\varepsilon
= N_\Lambda^{-1} \int_{-\frac{\Lambda}{2}}^{\frac{\Lambda}{2}} d x
\left[ \frac{\hbar^2 A_x^2}{2m} +
 \frac{m n}{2} \left(\frac{\hbar \Theta_x}{m}\right)^2 
+ \frac{g n^2}{2} \right] \, ,
\label{eq:cn_en_int}
\end{equation}
where the prefactor accounts for the number of particles in each
layer, $N_\Lambda = \bar{n} \Lambda$. The evaluation of the integral
in the above expression can be simplified using Eqs.~\eqref{eq:GP_Sx},
\eqref{eq:GP_Ax}, \eqref{eq:cn_ns}, and~\eqref{eq:cn_roots}. The final
result is
\begin{equation}
\begin{split}
\frac{\varepsilon}{g\bar{n}} = {}&{} 
1 + \frac{\eta}{2} \left[ 3 \Gamma(m_e) + m_e - 2 \right] \\
{}&{} + \frac{\eta^2}{6} \left[ 3 \Gamma^2(m_e) 
- 2 (2 - m_e) \Gamma(m_e) + 1 - m_e \right] \, .
\end{split}
\label{eq:cn_en_prop}
\end{equation}
We have checked that the minimization of $\varepsilon$ with respect
to $m_e$ and $\eta$ at fixed average density $\bar{n}$ and
velocity $|\bar{v}| > c$ always gives a uniform state.
Hence, unlike in the case of superfluid helium~\cite{Pitaevskii1984} and of
Bose gases with finite-range interaction~\cite{Baym2012}, here there is no
spontaneous transition from a uniform to a layered structure when the fluid
velocity crosses the critical one (equal to $c$ in our case). For this reason
cnoidal waves should be regarded as (nonlinear) excited states of the system.

\section{Dynamic properties}
\label{sec:dyn_properties}
This section is devoted to the study of the quantities characterizing
the dynamic behavior of a cnoidal wave. We first derive the Bogoliubov
equations (Sec.~\ref{subsec:dyn_Bogo_eqs}). Then, in
Sec.~\ref{subsec:dyn_ex_spectrum} we compute and discuss the
excitation spectrum, whereas in Sec.~\ref{subsec:dyn_struc_fact} we
study the dynamic structure factor, its moments and the sum rules they
obey. We note here that the spectrum of cnoidal waves has been studied
by the mathematical physics community (see, e.g., Refs.~\cite{Rowlands1974,
Bottman2011,Gallay2015,Gustafson2017} and references therein), which mainly
addressed the problem of dynamic stability; our focus is different and
concerns the energetic instability on one side, and the relationship with
the phenomenon of supersolidity on the other side.

\subsection{Bogoliubov equations}
\label{subsec:dyn_Bogo_eqs}
We shall now use the Bogoliubov
approach~\cite{Castin_review,Pethick_Smith_book,Pitaevskii_Stringari_book}
to study small oscillations about the equilibrium configuration
derived in Sec.~\ref{subsec:cn_sol_der}.  
In the present context it is
convenient to describe the collective modes in terms of the
fluctuations of the density and the phase. To this aim we decompose
the total density and phase as $n(x) + \delta n(x,t)$ and
$\Theta(x) + \delta \Theta(x,t)$, respectively. At first order in
$\delta n$ and $\delta \Theta$
Eqs.~\eqref{eq:GP_td_phase}--\eqref{eq:GP_td_amp} become
\begin{subequations}
\label{eq:fl_eq}
\begin{align}
\delta n_t = {}&{} - \frac{\hbar}{m} \left( \Theta_x \delta n_x
+ \Theta_{xx} \delta n + n \delta \Theta_{xx}
+ n_x \delta \Theta_x \right) \, ,
\label{eq:fl_dens_eq} \\
\begin{split}
\delta \Theta_t = {}&{} \frac{\hbar}{4 m} \left( \frac{\delta n_{xx}}{n}
- \frac{n_x}{n^2} \, \delta n_x +
\frac{n_x^2 - n n_{xx}}{n^3} \, \delta n \right) \\
&{} -\frac{\hbar \Theta_x}{m} \delta \Theta_x -
\frac{g}{\hbar} \, \delta n \, .
\end{split}
\label{eq:fl_phase_eq}
\end{align}
\end{subequations}
We look for solutions oscillating in time of the form
\begin{subequations}
\label{eq:fl_ansatz}
\begin{align}
\delta n(x,t) &{} = \delta\tilde{n}(x) e^{- i \omega t}
+ \delta\tilde{n}^*(x) e^{i \omega t} \, ,
\label{eq:fl_dens_ansatz} \\
\delta \Theta(x,t) &{} = \delta\tilde{\Theta}(x) e^{- i \omega t}
+ \delta\tilde{\Theta}^*(x) e^{i \omega t} \, .
\label{eq:fl_phase_ansatz}
\end{align}
\end{subequations}
This turns Eqs.~\eqref{eq:fl_eq} into an eigenvalue problem, which
enables one to determine the frequency $\omega$ and the complex
amplitudes $\delta\tilde{n}$ and $\delta\tilde{\Theta}$. The latter obey
the normalization condition~\cite{Pitaevskii_Stringari_book}
\begin{equation}
i \int_{-\frac{\Lambda}{2}}^{\frac{\Lambda}{2}} dx 
\left[ \delta\tilde{n}^*(x) \delta\tilde{\Theta}(x) - 
\delta\tilde{\Theta}^*(x) \delta\tilde{n}(x) \right] = 1 \, .
\label{eq:fl_orthonorm} 
\end{equation}
For each solution $\delta\tilde{n}$ and $\delta\tilde{\Theta}$ with
frequency $\omega$ there exists another one, $\delta\tilde{n}^*$ and
$\delta\tilde{\Theta}^*$, having frequency $-\omega$~\cite{Castin_review}.
The integral of Eq.~\eqref{eq:fl_orthonorm} evaluates to $-1$ (instead of
$1$) for the latter solution. Both solutions correspond to the same
physical oscillation, as clear from the structure of Eqs.~\eqref{eq:fl_ansatz}.
In order to avoid this redundancy we shall only consider solutions having
positive norm. This choice is customary because, in a second-quantization
framework, it is naturally associated to the usual boson commutation relation.

\subsection{Excitation spectrum}
\label{subsec:dyn_ex_spectrum}
\begin{figure}
\includegraphics[scale=1]{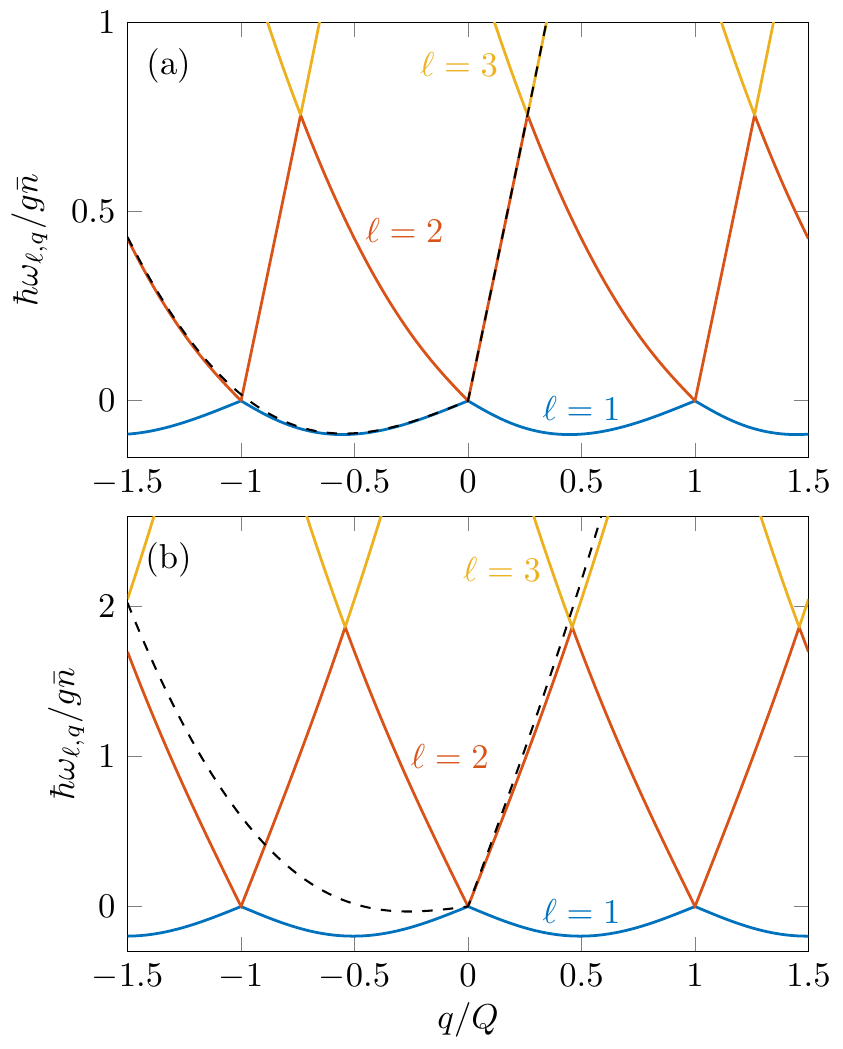}
\caption{Lowest three bands of the excitation spectrum as functions of
$q$. In panel (a) we take the same parameters $\eta = 0.6$ and $m_e = 0.5$
as the red dashed curve of Fig.~\ref{fig:cn_dens_prof}(a). In panel (b)
we choose $\eta = 2.0$ and a value $m_e = 0.8$ very close to
$m_e^\mathrm{max} = 0.826$ [see yellow dash-dotted curves of
Figs.~\ref{fig:cn_cont_sf_frac}(b)-(d)]. In both cases the sign of the
current density $\mathcal{J}$ is taken positive. The black dashed lines
show the spectrum of a uniform system having the same average density
$\bar{n}$ and velocity $\bar{v}/c = 1.18$ (a), $1.10$ (b) as the cnoidal
waves under consideration.}
\label{fig:fl_spectrum}
\end{figure}

The procedure for solving this eigenvalue problem is similar to that
employed in the previous works~\cite{Li2013,Martone2018}, and is
detailed in Appendix~\ref{sec:dyn_calculation}. Since the coefficients of
the linear coupled equations~\eqref{eq:fl_eq} are periodic in $x$, we
can look for solutions $\delta\tilde{n}$ and $\delta\tilde{\Theta}$ in the
form of Bloch waves~\cite{Ashcroft_Mermin_book}. They are given by a
plane wave, with wave vector $q$, times a periodic function with
period $\Lambda$ [see Eqs.~\eqref{eq:fl_amp}]. To any fixed value of
$q$ there correspond infinitely many solutions, with different
amplitudes and frequencies. This is at the origin of the band
structure exhibited by the Bogoliubov spectrum. This structure is
clearly visible in Fig.~\ref{fig:fl_spectrum}, where we plot the
lowest three bands of the spectrum of elementary excitations
of two given cnoidal-wave solutions. To distinguish between
the various Bogoliubov modes we make use of two subscripts,
the quasimomentum $q$ and the band index $\ell = 1,2,\ldots$.
The spectrum is periodic in $q$, with period $Q = 2 \pi / \Lambda$
equal to the wave vector of the density modulations. Each range
of values of $q$ enclosed between consecutive integer multiples
of $Q$ defines a Brillouin zone. Notice that the frequencies
$\omega_{\ell,q}$ are not invariant under inversion of $q$ into $- q$;
this reflects the fact that cnoidal-wave solutions do not enjoy parity
and time-reversal symmetry separately when the current $\mathcal{J}$
they carry is not zero. For the sake of comparison, in each panel of
Fig.~\ref{fig:fl_spectrum} we also plot (dashed curve) the spectrum of
a uniform Bose gas having the same average density $\bar{n}$ and
velocity $\bar{v}$ as the cnoidal wave considered in the panel.

The main feature of the spectra of Fig.~\ref{fig:fl_spectrum} is
that the two lowest bands ($\ell = 1,2$) are gapless and have linear
dispersion close to the edges of each Brillouin zone. The higher branch
($\ell = 2$) at small positive $q$ and the lower one ($\ell = 1$) at
negative $q$ are already present in a uniform system (dashed curve),
whereas the other two branches are specific of cnoidal
waves.\footnote{This actually holds only for positive $\bar{v}$; the
situation is reversed when $\bar{v} < 0$.} 

The presence of two gapless Goldstone modes is a feature expected
for the spectrum of a one-dimensional supersolid intended as a system
which breaks both $\mathrm{U}(1)$ and continuous translational invariance
(see, e.g., Ref.~\cite{Watanabe2012} for a detailed discussion).   
Such an increase of gapless modes has been indeed theoretically discussed
for the supersolid phase of solid Helium~\cite{Andreev1969}, of soft-core
Bose gases~\cite{Rica2007,Kunimi2012,Saccani2012,Macri2013}, of dipolar
Bose gases, as well as for the stripe phase in spin-orbit-coupled Bose
gases~\cite{Li2013,Liao2018}. It is under very active experimental
investigation for dipolar
gases~\cite{Natale2019,Tanzi2019b,Guo2019,Petter2020}.  

The two gapless bands of Fig.~\ref{fig:fl_spectrum}, as well as all the
upper bands of the excitation spectrum, have real frequency. This result
confirms that cnoidal waves in one-dimensional Bose gases with repulsive
contact interaction are dynamically stable, as pointed out in
Refs.~\cite{Rowlands1974,Bottman2011,Gallay2015,Gustafson2017}. However,
different from other supersolids considered in literature, here the frequency
of the lowest gapless band is negative, meaning that these waves are
energetically unstable. This agrees with the fact that they are excited
states of the system, as discussed in
Sec.~\ref{subsubsec:cn_energy_part}.\footnote{Similar negative-frequency
modes, although of discrete nature, have been found in soliton trains
trapped in ring geometries~\cite{Kanamoto2009}.} Such an instability
can lead to the decay of a cnoidal wave to a lower-energy state if one
applies an external perturbation. However, if this decay takes place over
sufficiently long timescales,
measurements of the dispersion relation based on Bragg spectroscopy
techniques would still be feasible. It is worth mentioning that the situation
is not very different from that of dipolar gases. Indeed, due to three-body
losses (energetic instability) the lifetime of the supersolid phase in those
systems is very short (few to tens milliseconds) but many measurements,
from phase coherence to collective excitations to Bragg spectroscopy,
have been performed~\cite{Tanzi2019a,Boettcher2019,Chomaz2019,
Natale2019,Tanzi2019b,Guo2019,Petter2020,Tanzi2019c,Ilzhoefer2019},
aiming at confirming the supersolid behavior. This phenomenon can be
referred to as ``transient supersolidity''.

A practical consequence of the structure of the excitation spectrum we
just discussed is that our system cannot flow around an external obstacle
without dissipating energy. It well known that for non-uniform systems
(such as Bose-Einstein condensates in optical lattices and supersolids)
the question of drag in the presence of an external obstacle is a different
issue from superfluidity. The latter corresponds to a dissipationless flow
of particles of the fluid through the fluid itself~\cite{Svistunov_book}
(see also the discussion in Sec. I. B. of Ref.~\cite{Boninsegni_review}),
testified by the existence of a finite superfluid fraction, which for cnoidal
waves is given by Eq.~\eqref{eq:cn_sf_frac}. The occurrence of drag in a
supersolid flowing past an external body was first pointed out by Pomeau and
Rica in Ref.~\cite{Pomeau1994}. Subsequently, Ref.~\cite{Martone2018}
computed the drag force experienced by a spin-orbit-coupled Bose-Einstein
condensate in the supersolid stripe phase moving through a pointlike impurity,
showing that energy dissipation occurs at any condensate speed. In these examples
drag occurs because the lowest-lying bands of the Bogoliubov spectrum have
vanishing frequency at the edges of each Brillouin zone, yielding a zero
Landau critical velocity. In the case of cnoidal waves, the energetic instability
can make the effects of the drag more dramatic because the interaction with
an external body can populate the negative-frequency modes.

The nature of the two gapless bands can be understood looking at the
limit where their frequencies vanish. This happens when $q$ lies on an
edge of a Brillouin zone. Setting $\delta n_t = 0$ and
$\delta \Theta_t = 0$ in Eqs.~\eqref{eq:fl_eq} one finds two kinds of
solutions. The first one is $\delta n = 0$ and $\delta \Theta$ equal to a
constant; it corresponds to an infinitesimal $\mathrm{U}(1)$
transformation of the phase of the condensate wave function. The
second solution is $\delta n = n_x \delta x_0$ and
$\delta \Theta = \Theta_x \delta x_0$, which performs a translation of the wave
function by an infinitesimal displacement $\delta x_0$. This finding further
reinforces the idea that the appearance of these modes is a result of the
spontaneous breaking of $\mathrm{U}(1)$ and continuous translational symmetry.

For modes with nonzero frequency one can still distinguish between a
crystal and phase character. The former involves small oscillations of
the density peaks about their equilibrium positions; the latter
features a superfluid current of particles tunneling from one peak to
another~\cite{Saccani2012,Macri2013,Natale2019,Guo2019}. However,
hybridization can occur and both characters can be present in a single
mode when $\omega_{\ell, q} \neq 0$. In cnoidal waves with small
$m_e$, the $\ell = 1$ branch at $q \lesssim 0$ and the $\ell = 2$ one
at $q \gtrsim 0$ have a dominant phase character, which is explained
by their closeness to the corresponding modes of a uniform gas [dashed
curve of Fig.~\ref{fig:fl_spectrum}(a)]; conversely, the branches that do
not appear in uniform systems are mainly crystal modes. This change of
behavior when crossing $q = 0$ becomes less and less pronounced at
increasing $m_e$ because of stronger hybridization. When $m_e$ is
large [close to $1$ for $\eta \leq 1$ and to $m_e^\mathrm{max}$ for
$\eta > 1$, as in Fig.~\ref{fig:fl_spectrum}(b)] we find that the $\ell = 1$
branch becomes dominantly crystal-like for both positive and negative
$q$. In particular, in the $\eta \leq 1$ and $m_e \lesssim 1$ case the
frequency of this band is almost $0$ at all $q$, and in the $m_e \to 1$
limit it reduces to the zero-frequency mode of the excitation spectrum
of a dark soliton. For $\eta > 1$ and $m_e$ close to $m_e^\mathrm{max}$
the phase character of both gapless bands is further suppressed
because of the strong reduction of the superfluid fraction pointed out
in Sec.~\ref{subsubsec:cn_sf_frac}.

As we mentioned in Sec.~\ref{subsubsec:cn_dens_prof}, at
$\eta \leq 1$, $m_e$ can reach values close to unity, and in this
regime cnoidal waves can be regarded as chains of dark solitons. It is
proven in Appendix~\ref{sec:dyn_lower_branch} that in this case
the dispersion relation
of the lowest band has the following analytic expression:
\begin{equation}
\frac{\hbar\omega_{1,q}}{g \bar{n}}  = - 8\,  \eta^{3/2} 
\, e^{-\sqrt{\eta} \Lambda / \xi} 
|\sin({\pi q}/{Q})| \, .
\label{eq:fl_low_mode}
\end{equation}
We have checked that this expression reproduces very accurately the lower
branch of the spectrum in the regime where $m_e\to 1$ and
$\eta \ll - \frac{1}{2} \ln(1-m_e)$, see Appendix~\ref{sec:dyn_lower_branch}.

Finally, we examined the regions of the spectrum where two different
bands approach each other and tried to determine whether they cross or
not. Our numerical results suggest that there is no gap separating any
couple of adjacent bands, and thus we are in the presence of a phenomenon
of level crossing. Hence, the usual argument of gap opening because of Bragg
scattering at the boundary of the Brillouin zone~\cite{Ashcroft_Mermin_book}
does not seem to apply here, presumably because cnoidal waves do not scatter
linear excitations.

\subsection{Dynamic structure factor and sum rules}
\label{subsec:dyn_struc_fact}
The dynamic structure factor provides important information on the
dynamic behavior of the system. It is
given by
\begin{equation}
S(q,\omega) = \sum_{\ell=1}^{\infty} 
|\langle 0| \delta\rho_q| \ell \rangle|^2 
\delta(\hbar\omega - \hbar\omega_{\ell,q}) \, ,
\label{eq:fl_dsf}
\end{equation}
where the sum is extended over all the bands, and $\delta\rho_q$ is
the $q$-component of the density fluctuation operator. Its matrix
operator between the ground state $|0\rangle$ and the $\ell$-th
excited band $|\ell\rangle$ can be easily computed:
$\langle 0| \delta\rho_q| \ell \rangle = \int_{-\Lambda/2}^{\Lambda/2}
dx \, \delta \tilde{n}_{\ell,q}(x) e^{- i q x} = \Lambda
\delta\tilde{n}_{\ell,q,0}$,
where $\delta\tilde{n}_{\ell,q,0}$ is the $\nu=0$ coefficient in
the Bloch-wave expansion of Eq.~\eqref{eq:fl_dens_amp}.

\begin{figure}
\includegraphics[scale=1]{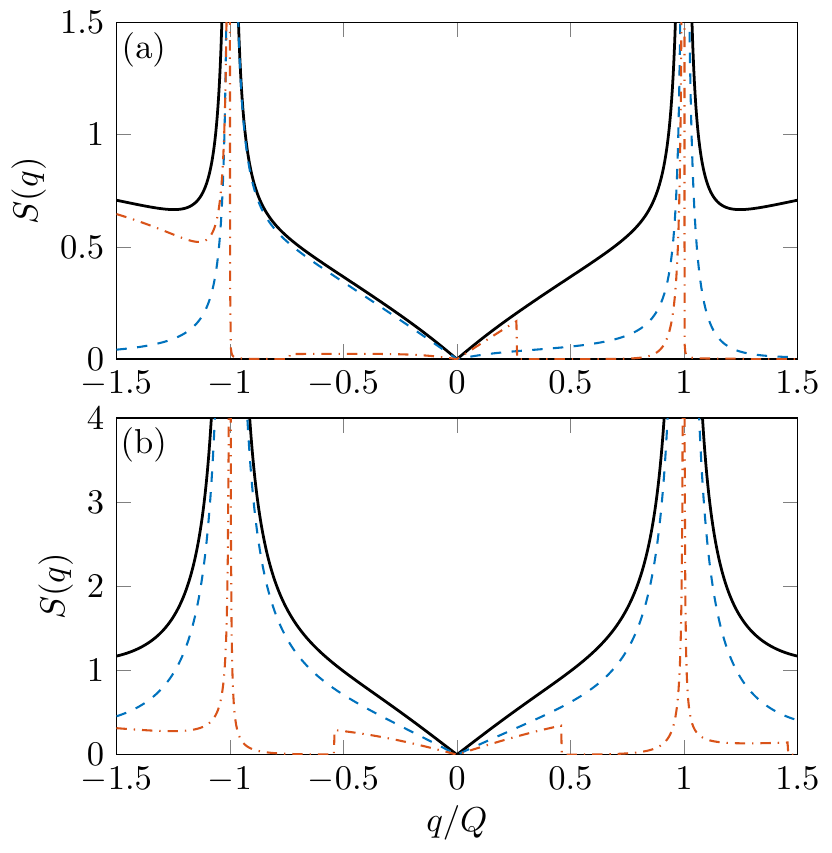}
\caption{Static structure factor as a function of $q$ (black solid
lines). The blue dashed and red dash-dotted curves show the
contribution of the lower and upper gapless band, respectively.
The cases (a) and (b) correspond to the same values of $m_e$
and $\eta$ than in Fig.~\ref{fig:fl_spectrum}: $\eta=0.6$,
$m_e=0.5$, and $\eta=2.0$, $m_e=0.8$, respectively.}
\label{fig:fl_ssf}
\end{figure}

For a given integer $p$ one defines the $p$-th moment of the dynamic
structure factor as~\cite{Pitaevskii_Stringari_book}
\begin{equation}
\begin{split}
m_p(q) &{} = \hbar^{p+1} 
\int_{-\infty}^{+\infty} d\omega \, \omega^p S(q,\omega) \\
&{} = \sum_{\ell=1}^{\infty} 
(\hbar\omega_{\ell,q})^p \, |\langle 0| \rho_q| \ell \rangle|^2 \, .
\end{split}
\label{eq:fl_mp}
\end{equation}
We first consider the $p=0$ moment $m_0(q) = N_\Lambda S(q)$,
where we have introduced the static structure factor
\begin{equation}
S(q) = N_\Lambda^{-1} \sum_{\ell=1}^{\infty} 
|\langle 0| \rho_q| \ell \rangle|^2 \, .
\label{eq:fl_ssf}
\end{equation}
We plot $S(q)$ in Fig.~\ref{fig:fl_ssf} for the same values of
the parameters as Fig.~\ref{fig:fl_spectrum}. We also plot the
contributions of the two gapless bands (sometimes referred
to as the ``strengths'' of $\delta\rho_q$). These contributions are
not symmetric under inversion of $q$ into $-q$ for the same reason the
spectra of Fig.~\ref{fig:fl_spectrum} are not; however, as was
shown in Ref.~\cite{Martone2012}, the full static structure factor is
symmetric as a consequence of its definition, regardless of the
properties of the ground state.

The contributions of the gapless bands to $S(q)$ are dominant
at small $|q|$. For a shallow cnoidal wave (small $m_e$)
the lower gapless band ($\ell=1$) exhausts $S(q)$ at $q < 0$ and the
upper one ($\ell=2$) at $q > 0$, as visible in Fig.~\ref{fig:fl_ssf}(a).
This behavior has the same explanation as that of the excitation
spectrum (see Sec.~\ref{subsec:dyn_ex_spectrum}), namely, it stems
from the closeness of these shallow waves to uniform gases. It is also
shared by all the moments $m_p(q)$ with $p \neq 0$. As $m_e$
increases, the strength of the lower band grows significantly and
eventually, at high $m_e$, it dominates $S(q)$ in a wide range of $q$
(both positive and negative), as shown in Fig.~\ref{fig:fl_ssf}(b).

Another remarkable feature is that the strengths of both gapless
bands [and consequently $S(q)$ itself] diverge when $q$ equals an
integer multiple of $Q$, i.e., at the edges of each Brillouin zone
(except at $q=0$). An analogous behavior occurs for the supersolid
phase of dipolar gases~\cite{Kora2019,RoccuzzoPhD}, as well as for
the stripe phase of spin-orbit-coupled Bose-Einstein
condensates~\cite{Li2013}, where, using sum-rule arguments, it was
shown that the existence of a nonzero crystalline order parameter
causes a $|q- Q|^{-1}$ divergence of $S(q)$ at the boundary of the
first Brillouin zone.

\begin{figure}
\includegraphics[scale=1]{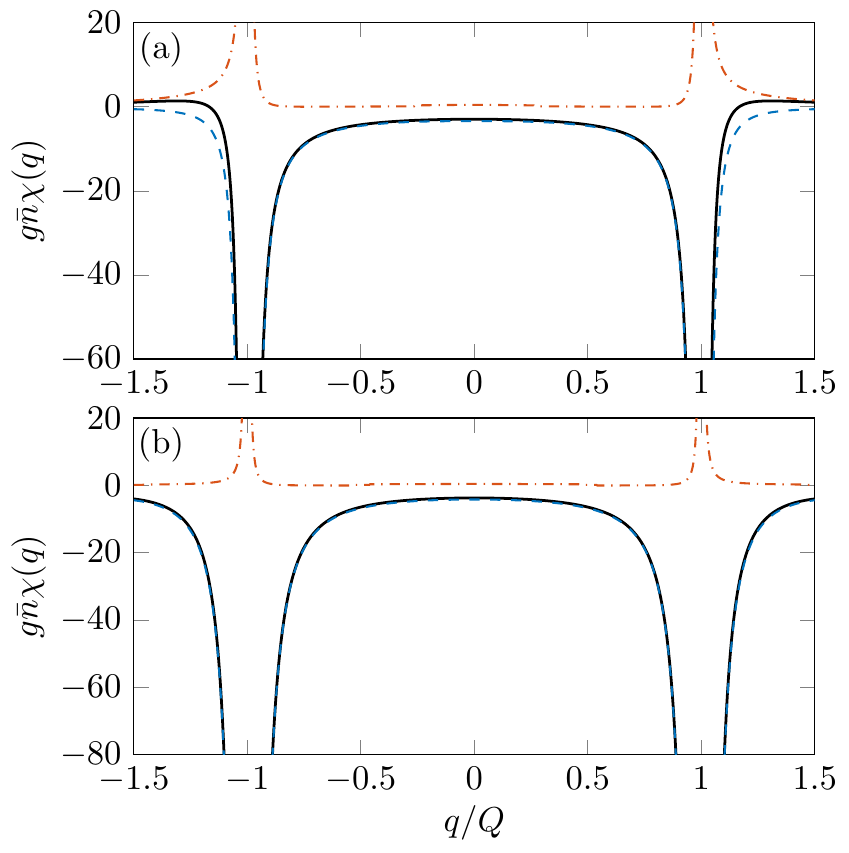}
\caption{Compressibility as a function of $q$ (black solid lines).
The blue dashed and red dash-dotted curves show the contribution
of the lower and upper gapless band, respectively.
The cases (a) and (b) correspond to the same values of $m_e$
and $\eta$ than in Fig.~\ref{fig:fl_spectrum}: $\eta=0.6$,
$m_e=0.5$, and $\eta=2.0$, $m_e=0.8$, respectively.}
\label{fig:fl_chi}
\end{figure}

The $p = -1$ moment is related to the compressibility $\chi(q)$ by
\begin{equation}
m_{-1}(q) + m_{-1}(-q) = N_\Lambda \chi(q) \, .
\label{eq:fl_chi}
\end{equation}
The behavior of the
compressibility, as well as that of the contributions of the two
gapless bands, is displayed in Fig.~\ref{fig:fl_chi}. Notice that
$\chi(q)$ is dominated by the lowest negative-frequency band for a
wide range of values about $q=0$, and is thus itself negative in this
range, revealing once more the presence of an energetic instability.
Like the static structure factor, it diverges at the edges of
the Brillouin zones except $q=0$, again in agreement with the findings
of Ref.~\cite{Li2013}. Interestingly, as illustrated in
Fig.~\ref{fig:fl_chi}, the negative divergence of the
total $\chi(q)$ is caused by the contribution of the $\ell = 1$ band,
whereas the $\ell = 2$ term is positively divergent. One can better
understand this aspect using a sum-rule argument. From the inequality
$\hbar\omega_{1,q} m_{-1}(q) \leq m_0(q)$, which holds even for negative
$\omega_{1,q}$, it follows that
\begin{equation}
\chi(q) \geq
\left( \frac{1}{\hbar\omega_{1,q}}
+ \frac{1}{\hbar\omega_{1,-q}} \right) S(q) \, .
\label{eq:fl_chi_ineq}
\end{equation}
The right-hand side of Eq.~\eqref{eq:fl_chi_ineq} approximates well
$\chi(q)$ when $q$ is close to $\pm Q$ and the static structure factor
is exhausted by the $\ell = 1$ term; it negatively diverges as
$q \to \pm Q$ because the prefactor of $S(q)$ is negative. This
divergence is however mitigated by the contributions of the
positive-frequency modes, which is again consistent with the
inequality~\eqref{eq:fl_chi_ineq}. 

Finally, we have checked that the $p=1$ moment satisfies the $f$-sum
rule $m_1(q) + m_1(-q) = N_\Lambda \hbar^2 q^2 /
m$~\cite{Pitaevskii_Stringari_book}.
Different from the sum rules discussed previously, for large $m_e$ and
small $|q|$ the $f$-sum rule is dominated mainly by the upper gapless
band. This is because the lower band, despite having bigger strength,
has much smaller frequency (in absolute value) than the upper one in
this regime.

\section{Conclusion}
\label{sec:conclusion}
We have studied several relevant features of an ultracold Bose gas in
a cnoidal-wave state. The equilibrium wave function is characterized
by periodic spatial density modulations described in terms of Jacobi's
elliptic functions. Cnoidal waves spontaneously break both
$\mathrm{U}(1)$ and continuous translational symmetry, thus exhibiting
typical supersolid features.  Besides, as argued by 
Leggett~\cite{Leggett1970,Leggett1998}, their superfluid fraction is
depleted even at zero temperature, and gets smaller and smaller as the
contrast of the fringes increases. A further signature of supersolidity
is represented by the behavior of the excitation spectrum, featuring a
band structure with two gapless bands. The latter exhibit a mixed
phase and crystal character. The presence of a crystalline structure
causes the divergence of the static structure factor and the
compressibility at the edges of the Brillouin zones.

The configurations studied in the present work could remind the Abrikosov
lattice in a two-dimensional system. However, in the latter case one has a
redundancy among the Nambu-Golstone bosons~\cite{Watanabe2013},
leading to a spectrum with a single quadratic mode. Interestingly, the same
does not occur in our case and indeed two (linear) modes are present in the
spectrum, related to the $\mathrm{U}(1)$ and translational symmetries.

Our results open new perspectives for the study of supersolidity in
ultracold atomic gases. Cnoidal waves are excited states that could
be realized, for instance, moving an obstacle into the gas at an
appropriate speed~\cite{Onofrio2000,Pavloff2002,Engels2007,
Leszczyszyn2009,Dries2010,Kamchatnov2012}. The density modulations
can be probed either \textit{in situ} or after time of flight. The
excitation spectrum and the dynamic structure factor can then be
accessed using two-photon Bragg spectroscopy.

The spectrum of a cnoidal wave demonstrates dynamic stability (all
the eigenvalues are real) and, more importantly, energetic instability 
(some of the eigenvalues are negative). This aspect may be relevant in
the context of analog gravity: it has been shown~\cite{Leboeuf2001,
Kamchatnov2012} that in some circumstances, an
obstacle moving at supersonic speed in a Bose-Einstein condensate may
give rise to an upstream cnoidal wave and a downstream supersonic flat
density pattern, both stationary in the reference frame of the
obstacle. It would then be of great interest to study the analogous
Hawking radiation in this realistic, and experimentally relevant
setting, where negative-norm modes exist on both sides of the acoustic
pseudo-horizon~\cite{Balbinot2013}.

We finally note that various types of cnoidal waves have already been
experimentally realized in the framework of nonlinear optics, both for
repulsive and attractive interaction, in the out-of-equilibrium context
of nonlinear whispering-gallery mode resonators~\cite{Coillet2013,
Herr2014,Pfeifle2015,Cole2017}, but also in two-dimensional
photo-refractive media~\cite{Petter2003,Desyatnikov2005} and in
optical fibers~\cite{Xu2020}. We believe it should be possible to study
supersolidity phenomena in such settings, even in the presence of
dissipation, as has been done for superfluidity in non-equilibrium
systems of condensed microcavity polaritons~\cite{Amo2009}.

\begin{acknowledgments}
We thank F. Dalfovo, A. Fabbri, D. Frantzeskakis, A. M. Kamchatnov,
Y. V. Kartashov, P. Kevrekidis, T. Paul, P. Pedri, L. P. Pitaevskii,
S. Stringari, and G. Theocharis for fruitful discussions. 
The research leading to these results has received funding from the
European Research Council under European Community's Seventh
Framework Programme (FP7/2007-2013 Grant Agreement No. 341197).
We acknowledge the support from Provincia Autonoma di Trento,
and the Italian MIUR under the PRIN2017 project CEnTraL.
\end{acknowledgments}

\appendix
\section{Limiting cases}
\label{sec:cn_lim_cases}
As discussed in Ref.~\cite{Tsuzuki1971}, the cnoidal-wave solution
admits several important limiting cases. In this appendix we shall
focus on the uniform and linear-wave limit at small $m_e$, as well
as on the dark-soliton limit corresponding to $m_e \to 1$.

\subsection{Uniform limit}
\label{subsec:cn_unif}
Let us first look at the $m_e = 0$ case. In this situation the
amplitude of the density oscillations vanishes. Besides, the
velocity~\eqref{eq:cn_avg_vel} simplifies to
$\bar{v} = \pm c \sqrt{1 + \eta}$. This yields $\eta = \eta_0$ with
$\eta_0 = \bar{v}^2 / c^2 - 1$, and implies that the flow is supersonic,
i.e., $|\bar{v}| \geq c$ (recall that $\eta > 0$ by definition).
The chemical potential~\eqref{eq:cn_ns_mu} takes the standard form
$\mu = g \bar{n} + m \bar{v}^2/2$ (the same happens in the linear-wave
and dark-soliton limits discussed below).

\subsection{Linear-wave limit}
\label{subsec:cn_lin}
At first order in $m_e$ one can approximate
$\Gamma(m_e) - 1 \simeq - m_e / 2$ and replace the $\sn$ function with
an ordinary sine in Eq.~\eqref{eq:cn_mod_dens}. As a consequence, in this
limit (nonlinear) cnoidal waves reduce to (linear) sinusoidal waves,
\begin{align}
\frac{n(x)}{\bar{n}} &{} \simeq 1 - \frac{\eta_0 m_e}{2} 
\cos \left( Q_0 x \right) \, ,
\label{eq:cn_dens_Bogo} \\
\Theta(x) &{} \simeq \frac{m \bar{v} x}{\hbar}
\pm \frac{\sqrt{\eta_0 (1 + \eta_0)} m_e}{4} \sin \left( Q_0 x \right) \, ,
\label{eq:cn_phase_Bogo}
\end{align}
where $Q_0 = 2 \sqrt{\eta_0} / \xi = 2 m \sqrt{\bar{v}^2 - c^2} / \hbar$.

It is interesting to study these waves in a frame where they travel with
a given phase velocity $V$. In such a frame, at every point in space the density
and the velocity field oscillate in time with frequency $\Omega_0 = Q_0 V$.
Let us set $\bar{v} = \bar{v}' - V$, where $\bar{v}'$ is the average velocity
in the new frame. One can invert the definition of $Q_0$ given just above
to express $V$, and hence $\Omega_0$, as a function of $\bar{v}'$
and $Q_0$. One finds the two values
\begin{equation}
\hbar\Omega_0(Q_0) = \hbar Q_0 \bar{v}' \mp \sqrt{\frac{\hbar^2 Q_0^2}{2 m} 
\left(\frac{\hbar^2 Q_0^2}{2 m} + 2 g \bar{n} \right) } \, .
\label{eq:cn_freq_Bogo}
\end{equation}
These are the two branches of the Bogoliubov spectrum of the uniform
condensate discussed in the previous section (which flows with velocity
$\bar{v}'$ in the new frame). Besides, the ratio between the amplitudes
of the phase and density modulations is $\mp \sqrt{(1+\eta_0)/(4\eta_0)}
= \mp m |\Omega_0 - Q_0 \bar{v}'| / \hbar Q_0^2$, in agreement with the
prediction of the Bogoliubov theory~\cite{Pitaevskii_Stringari_book}.
We thus conclude that in the small-$m_e$ limit traveling cnoidal waves
describe standard Bogoliubov modes.

\subsection{Dark-soliton limit}
\label{subsec:cn_sol}
In the $m_e \to 1$ limit one has $\Gamma(m_e) \to 0$ and the $\sn$
function approaches the hyperbolic tangent.  The
velocity~\eqref{eq:cn_avg_vel} becomes $\bar{v} = \pm c \sqrt{1 - \eta}$.
This gives $\eta = 1 - \bar{v}^2 / c^2$, together with the condition of
subsonic flow $|\bar{v}| \leq c$. Thus, the density and phase turn into
those of a dark soliton~\cite{Pethick_Smith_book,Pitaevskii_Stringari_book},
written in the frame where the latter is at rest and the background
has velocity $\bar{v}$:
\begin{align}
\frac{n(x)}{\bar{n}} = {}&{} 1
- \frac{\cos^2\theta}{\cosh^2 \left( \cos\theta \, x/\xi \right)} \, ,
\label{eq:cn_dens_sol_ds} \\
\Theta(x) = {}&{} \frac{m \bar{v} x}{\hbar}
+ \arctan \left[ \frac{\tanh(\cos\theta \, x/\xi)}{\tan\theta} \right] \, .
\label{eq:cn_phase_sol_ds}
\end{align}
Here we have set $\sin\theta = \bar{v}/c$ and $\cos\theta = \sqrt{\eta}$,
with $\theta \in [-\pi/2,\pi/2]$. In particular, when $\bar{v} = 0$ one obtains
a black soliton, whose density vanishes at the center.

\section{Calculation of the Bogoliubov frequencies and amplitudes}
\label{sec:dyn_calculation}
In this appendix we present the method we have used for solving the
Bogoliubov equations~\eqref{eq:fl_eq}. Using the
Ansatz~\eqref{eq:fl_ansatz}, and equating the coefficients of $e^{- i
\omega t}$ and $e^{i \omega t}$ on both sides, they can be cast in
matrix form as follows:
\begin{equation}
\begin{pmatrix}
\mathcal{B}^{(nn)} & \mathcal{B}^{(n\Theta)} \\
\mathcal{B}^{(\Theta n)} & \mathcal{B}^{(\Theta\Theta)}
\end{pmatrix}
\begin{pmatrix}
\delta \tilde{n} \\
\delta \tilde{\Theta}
\end{pmatrix}
=
\omega
\begin{pmatrix}
\delta \tilde{n} \\
\delta \tilde{\Theta}
\end{pmatrix}
\, .
\label{eq:fl_eig_probl}
\end{equation}
Here we have defined the four operators
\begin{subequations}
\label{eq:fl_eig_op}
\begin{align}
\mathcal{B}^{(nn)} &{} = - \frac{i \hbar}{m}
\left( \Theta_x \frac{d}{d x} + \Theta_{xx} \right) \, ,
\label{eq:fl_eig_op_nn} \\
\mathcal{B}^{(n\Theta)} &{} = - \frac{i \hbar}{m}
\left( n \frac{d^2}{d x^2} + n_x \frac{d}{d x} \right) \, ,
\label{eq:fl_eig_op_nS} \\
\mathcal{B}^{(\Theta n)} &{} = \frac{i \hbar}{4 m}
\left( n^{-1} \frac{d^2}{d x^2} - \frac{n_x}{n^2} \frac{d}{d x}
+ \frac{n_x^2 - n n_{xx}}{n^3} \right)
- \frac{i g}{\hbar} \, ,
\label{eq:fl_eig_op_Sn} \\
\mathcal{B}^{(\Theta\Theta)}
&{} = -\frac{i \hbar \Theta_x}{m} \frac{d}{d x} \, .
\label{eq:fl_eig_op_SS}
\end{align}
\end{subequations}
Because of the periodicity of the coefficients entering these
operators, the solutions of Eq.~\eqref{eq:fl_eig_probl} can be
expressed as Bloch waves,
\begin{subequations}
\label{eq:fl_amp}
\begin{align}
\delta \tilde{n}_{\ell,q}(x) &{} =
e^{i q x} \sum_{\nu \in \mathbb{Z}} \delta \tilde{n}_{\ell,q,\nu}
e^{i \nu Q x} \, ,
\label{eq:fl_dens_amp} \\
\delta \tilde{\Theta}_{\ell,q}(x) &{} = e^{i q x} \sum_{\nu \in \mathbb{Z}}
\delta \tilde{\Theta}_{\ell,q,\nu} e^{i \nu Q x} \, .
\label{eq:fl_phase_amp}
\end{align}
\end{subequations}
Here $q$ denotes the quasimomentum, $\delta \tilde{n}_{\ell,q,\nu}$
and $\delta \tilde{\Theta}_{\ell,q,\nu}$
are Fourier expansion coefficient, and the sums run over all integer
$\nu$. As explained in Sec.~\ref{subsec:dyn_ex_spectrum}, the band
index $\ell$ is needed to distinguish between different solutions at
fixed $q$.

We now derive a set of equations that allow one to calculate the expansion
coefficients $\delta \tilde{n}_{\ell,q,\nu}$ and $\delta
\tilde{\Theta}_{\ell,q,\nu}$ for each Bogoliubov mode, as well as the
corresponding frequency $\omega_{\ell,q}$. For this purpose, we need
the Fourier expansions of the coefficients of
operators~\eqref{eq:fl_eig_op}. This analysis can be simplified
recalling that $-n_x/n^2 = (n^{-1})_x$, $\Theta_x = m \mathcal{J}
\hbar^{-1} n^{-1}$, $\Theta_{xx} = m \mathcal{J} \hbar^{-1} (n^{-1})_x$,
and
\[
\begin{split}
\frac{n_x^2 - n n_{xx}}{n^3} = {}&{}
\frac{1}{2} \left[ (n^{-1})_{xx} - \frac{n_{xx}}{n^2} \right] \\
= {}&{} \frac{(n^{-1})_{xx}}{2} - \frac{3 m g}{\hbar^2} +
\frac{4 m \mu}{\hbar^2 n} - \frac{2 m \mathcal{E}}{\hbar^2 n^2} \, .
\end{split}
\]
The expressions of $\mu$, $\mathcal{E}$, and $\mathcal{J}$ are
provided in Eqs.~\eqref{eq:cn_ns}. Thus, we only need to determine
the expansion of the density, $n(x) = \sum_{\nu \in \mathbb{Z}}
\tilde{n}_{\nu}^{(1)} e^{i \nu Q x}$, of its inverse
$n^{-1}(x) = \sum_{\nu \in \mathbb{Z}} \tilde{n}_{\nu}^{(-1)}
e^{i \nu Q x}$, and of the squared inverse $n^{-2}(x) =
\sum_{\nu \in \mathbb{Z}} \tilde{n}_{\nu}^{(-2)} e^{i \nu Q x}$.
Fourier coefficients of powers and rational fractions of
Jacobi's elliptic functions have been widely studied in
literature~\cite{Whittaker_Watson_book,Langebartel1980}.
For $\nu = 0$ one finds
\begin{subequations}
\label{eq:fl_coeff_m0}
\begin{align}
\tilde{n}_0^{(1)} &{} = \bar{n} \, ,
\label{eq:fl_coeff_m0_p1} \\
\tilde{n}_0^{(-1)} &{} = (f_s \bar{n})^{-1} \, ,
\label{eq:fl_coeff_m0_m1} \\
\begin{split}
\tilde{n}_0^{(-2)} &{} = \frac{(n_3-n_1)\Gamma(m_e) - n_3}{2 n_1 n_2 n_3} \\
&{} \phantom{={}} + \frac{n_1 n_2 + n_2 n_3 + n_3 n_1}{2 n_1 n_2 n_3}
\, (f_s \bar{n})^{-1} \, ,
\end{split}
\label{eq:fl_coeff_m0_m2}
\end{align}
\end{subequations}
where $f_s$ is the superfluid fraction defined in
Eq.~\eqref{eq:cn_sf_frac}. Instead, the $\nu \neq 0$
coefficients read
\begin{subequations}
\label{eq:fl_coeff_m}
\begin{align}
\tilde{n}_{\nu}^{(1)} &{} = - \frac{\pi^2 (n_3-n_1)}{2 K^2(m_e)}
\frac{\nu}{\sinh(\nu w)} \, ,
\label{eq:fl_coeff_m_p1} \\
\tilde{n}_{\nu}^{(-1)} &{} = \frac{\pi}{2 K(m_e)}
\sqrt{\frac{n_3-n_1}{n_1 n_2 n_3}}
\frac{\sinh(\nu w_0)}{\sinh(\nu w)} \, ,
\label{eq:fl_coeff_m_m1} \\
\tilde{n}_{\nu}^{(-2)} &{} =
\frac{(n_1 n_2 + n_2 n_3 + n_3 n_1) \tilde{n}_{\nu}^{(-1)} -
\cosh(\nu w_0) \tilde{n}_{\nu}^{(1)}}{2 n_1 n_2 n_3} \, .
\label{eq:fl_coeff_m_m2}
\end{align}
\end{subequations}
Here we have defined $w = \pi K(1-m_e)/K(m_e)$ and
$w_0 = \pi [K(1-m_e) - K_0]/K(m_e)$, where $0 < K_0 < K(1-m_e)$ is
solution of the equation $\cn(K_0,1-m_e) = \sqrt{(n_2-n_1)/n_2}$.

We now insert the Bloch-wave Ansatz~\eqref{eq:fl_amp} into the coupled
equations~\eqref{eq:fl_eig_probl}, and use the above results to expand
the coefficients of the operators~\eqref{eq:fl_eig_op} in Fourier
series. Then, we equate the terms on the two sides of the resulting
equations that oscillate in space with the same wave vector. This
yields an infinite set of coupled algebraic equations involving the
expansion coefficients $\delta \tilde{n}_{\ell,q,\nu}$ and $\delta
\tilde{\Theta}_{\ell,q,\nu}$, as well as the corresponding excitation
frequencies $\omega_{\ell,q}$. This set can be written in a compact
form by defining the two infinite-dimensional column vectors
\begin{align*}
\delta\tilde{\mathsf{n}}_{\ell,q}
&{} = ( \cdots \, \delta\tilde{n}_{\ell,q,\nu-1} \,\,\,
\delta\tilde{n}_{\ell,q,\nu} \,\,\, \delta\tilde{n}_{\ell,q,\nu+1} \,
\cdots)^T \, ,
\\
\delta\tilde{\mathsf{\Theta}}_{\ell,q}
&{} = ( \cdots \, \delta\tilde{\Theta}_{\ell,q,\nu-1} \,\,\,
\delta\tilde{\Theta}_{\ell,q,\nu} \,\,\,
\delta\tilde{\Theta}_{\ell,q,\nu+1} \, \cdots)^T \, .
\end{align*}
The normalization condition for $\delta\tilde{\mathsf{n}}_{\ell,q}$
and $\delta\tilde{\mathsf{\Theta}}_{\ell,q}$ follows from
Eq.~\eqref{eq:fl_orthonorm} and reads $i \Lambda
\left(\delta\tilde{\mathsf{n}}_{\ell,q}^\dagger
\delta\tilde{\mathsf{\Theta}}_{\ell,q} -
\delta\tilde{\mathsf{\Theta}}_{\ell,q}^\dagger
\delta\tilde{\mathsf{n}}_{\ell,q}\right) = 1$. The above procedure
leads to the eigenvalue equation
\begin{equation}
\begin{pmatrix}
\mathsf{B}^{(nn)}(q) & \mathsf{B}^{(n\Theta)}(q) \\
\mathsf{B}^{(\Theta n)}(q) & \mathsf{B}^{(\Theta\Theta)}(q)
\end{pmatrix}
\begin{pmatrix}
\delta\tilde{\mathsf{n}}_{\ell,q} \\
\delta\tilde{\mathsf{\Theta}}_{\ell,q}
\end{pmatrix}
=
\omega_{\ell,q}
\begin{pmatrix}
\delta\tilde{\mathsf{n}}_{\ell,q} \\
\delta\tilde{\mathsf{\Theta}}_{\ell,q}
\end{pmatrix}
\, ,
\label{eq:fl_eig_probl_mat}
\end{equation}
where the $\mathsf{B}$'s are infinite-dimensional matrices with entries
\begin{subequations}
\label{eq:fl_eig_mat}
\begin{align}
\mathsf{B}^{(nn)}_{\nu_1 \nu_2}(q) &{} =
\mathcal{J} \tilde{n}_{\nu_1-\nu_2}^{(-1)} (q + \nu_1 Q) \, ,
\label{eq:fl_eig_mat_nn} \\
\mathsf{B}^{(n\Theta)}_{\nu_1 \nu_2}(q) &{} =
\frac{i \hbar}{m} \tilde{n}_{\nu_1-\nu_2}^{(1)} (q + \nu_1 Q)
(q + \nu_2 Q) \, ,
\label{eq:fl_eig_mat_nS} \\
\begin{split}
\mathsf{B}^{(\Theta n)}_{\nu_1 \nu_2}(q) &{} = - \frac{i \hbar}{4 m}
\tilde{n}_{\nu_1-\nu_2}^{(-1)} \\
&{} \phantom{={}} \times
\left[ q^2 + (\nu_1+\nu_2) Q q +
\frac{(\nu_1^2+\nu_2^2)Q^2}{2} \right] \\
&{} \phantom{={}} - \frac{7 i g}{4 \hbar} \delta_{\nu_1,\nu_2}
+ \frac{i \mu}{\hbar} \, \tilde{n}_{\nu_1-\nu_2}^{(-1)}
- \frac{i \mathcal{E}}{2 \hbar} \, \tilde{n}_{\nu_1-\nu_2}^{(-2)} \, ,
\end{split}
\label{eq:fl_eig_mat_Sn} \\
\mathsf{B}^{(\Theta\Theta)}_{\nu_1 \nu_2}(q) &{} = \mathcal{J}
\tilde{n}_{\nu_1-\nu_2}^{(-1)} (q + \nu_2 Q) \, .
\label{eq:fl_eig_mat_SS}
\end{align}
\end{subequations}
Numerically solving Eq.~\eqref{eq:fl_eig_probl_mat} one recovers all
the results of Sec.~\ref{sec:dyn_properties}. Of course, in order to
reduce the problem to a finite-dimensional one it is necessary to fix
a cutoff $\nu_{\mathrm{max}}$, and truncate all the above Fourier
expansions retaining only the terms with $-\nu_{\mathrm{max}} \leq
\nu \leq \nu_{\mathrm{max}}$. The choice of the best value of
$\nu_{\mathrm{max}}$ depends on $m_e$ and $\eta$.  At fixed $\eta$
and small $m_e$, where cnoidal waves do not significantly deviate from
linear waves, taking $\nu_{\mathrm{max}}$ equal to $5$ or $6$ can
be sufficient to achieve good accuracy in the results. Conversely,
increasing $m_e$ one needs larger and larger values of
$\nu_{\mathrm{max}}$. These can even exceed $100$ when the cnoidal
wave is close to the soliton limit (for $\eta \leq 1$) or its contrast
is close to $1$ (for $\eta > 1$).

\section{The lower excitation branch of a train of dark solitons}
\label{sec:dyn_lower_branch}
In the regime where the period $\Lambda$ of the cnoidal wave is large
compared to the width $\xi/\sqrt{\eta}$ of a dark soliton, the cnoidal
wave can be considered as a train of regularly spaced identical
solitons. From Eq.~\eqref{eq:cn_wavelength} this occurs when $K(m_e) \gg
1$, i.e., when $m_e$ is close to unity. Since solitons are essentially
classical objects, it is natural to expect that, in this regime, the
lowest branch of the spectrum should be described as an excitation of
an array of classical particles connected by springs. Denoting by
$\Omega$ the resonant angular frequency associated to these springs,
the corresponding spectrum is of the form~\cite{Ashcroft_Mermin_book}
\begin{equation}
\omega_{1,q}=-2\, \Omega\, |\sin(q \Lambda/2)| \, .
\label{eq:lb_freq}
\end{equation}
The value of $\Omega$ depends on the interaction between two solitons
and on their inertial mass. It can be determined by means of Manton's
method~\cite{Manton1979,Kevrekidis2004} as explained now.

For studying the interaction between two solitons, one considers a
configuration where the solitons are stationary, in a background with
subsonic velocity $\bar{v}$ and otherwise uniform density $\bar{n}$.
It is convenient
to single out the velocity of the background and to write
$\psi(x,t)=\phi(x,t)\exp(i k x)$ where $k=m \bar{v}/\hbar$. Then, one
has in~\eqref{eq:GP} $\mu=\hbar^2 k^2/ 2 m + g \bar{n}$ and $\phi$ is
solution of:
\begin{equation}
i \hbar \phi_t = - \frac{\hbar^2}{2m} \phi_{xx} -i \frac{\hbar^2 k}{m} \phi_x 
+  g (|\phi|^2  - \bar{n}) \phi \, .
\label{eq:lb_gp}
\end{equation}
An Ansatz describing two identical stationary solitons separated by a
distance $\Delta$ is of the form
\begin{equation}
\phi(x)=\sqrt{\bar{n}} \, \Phi(x) \Phi(x-\Delta) \, ,
\label{eq:lb_ansatz}
\end{equation}
where
\begin{equation}
\Phi(x) =
\left[\cos\theta\tanh(\cos\theta \, x/\xi) - i \sin\theta\right] \, ,
\label{eq:lb_ds}
\end{equation}
with $\theta \in [-\pi/2,\pi/2]$.  $\sqrt{\bar{n}} \, \Phi(x)$
describes a stationary isolated soliton, solution of
Eq.~\eqref{eq:lb_gp}. The soliton is stationary because its velocity
$V=-c \sin\theta$ is exactly opposed to the velocity $\bar{v}=c
\sin\theta$ of the background. Notice that, up to a global phase
factor, one has $\sqrt{\bar{n}} \, \Phi(x) \exp(i k x)= \sqrt{n} \,
e^{i \Theta}$, with $n(x)$ and $\Theta(x)$ given by
Eqs.~\eqref{eq:cn_dens_sol_ds} and~\eqref{eq:cn_phase_sol_ds},
respectively. As regards the two-soliton case, of course the
Ansatz~\eqref{eq:lb_ansatz} is not an exact solution of the
Gross-Pitaevskii equation~\eqref{eq:lb_gp}, but it is expected to be a
reasonable approximation if\footnote{We will argue in the end of this
appendix that the regime of validity of the
Ansatz~\eqref{eq:lb_ansatz} needs to be defined more carefully.}
$\Delta \, \cos\theta \gg \xi$.

The Lagrangian density associated to the
Gross-Pitaevskii equation~\eqref{eq:lb_gp} is
\begin{equation}
\begin{split}
\mathcal{L}=
&  \frac{i \hbar}{2}(\phi^* \phi_t-\phi^*_t\phi) (1-\bar{n}|\phi|^{-2})
- \frac{\hbar^2}{2 m}|\phi_x|^2 \\
& -\frac{i \hbar^2 k}{2 m}( \phi_x^*\phi -\phi^* \phi_x)
- \frac{g}{2}(|\phi|^2-\bar{n})^2 \, .
\end{split}
\label{eq:lb_lagr}
\end{equation}
Note the unfamiliar multiplicative term $(1-\bar{n}|\phi|^{-2})$ in
the first term of the above expression. It corresponds to adding to
the usual Lagrangian density a total derivative which does not affect
the form of the Gross-Pitaevskii equation~\eqref{eq:lb_gp}, but yields
the correct physical momentum of a soliton~\cite{Ishikawa1980,
Shevchenko1988,Barashenkov1993,Barashenkov1994,Pitaevskii_Stringari_book},
namely
\begin{equation}
P=\hbar \, \bar{n} \left[\pi + 2\theta +\sin(2 \theta)\right] \, ,
\label{eq:lb_ds_mom}
\end{equation}
for a soliton of type~\eqref{eq:lb_ds}.

Considering two points $a$ and $b$ located around the soliton centered
at $\Delta$ ($a<\Delta<b$), one has
\begin{equation}
\frac{d}{d t}\int_a^b \!dx\, \mathcal{P}(x,t)=
\mathcal{T}(a,t)-\mathcal{T}(b,t) \, ,
\label{eq:lb_mom_der}
\end{equation}
where $\mathcal{P}=\frac{i \hbar}{2}(\phi \phi_x^*-\phi^* \phi_x)
(1-\bar{n} |\phi|^{-2})$ is the momentum density and
\begin{equation}
\begin{split}
\mathcal{T} =
{}&{} \frac{i \hbar}{2}(\phi^* \phi_t-\phi\phi^*_t) (1-\bar{n}|\phi|^{-2}) \\
{}&{} + \frac{\hbar^2}{2 m}|\phi_x|^2 - \frac{g}{2}(|\phi|^2-\bar{n})^2
\end{split}
\label{eq:lb_stress}
\end{equation}
the stress tensor, both associated to the Lagrangian
density~\eqref{eq:lb_lagr}.  If $\Delta-a$ and $b-\Delta$ are large
compared to $\xi / \cos\theta$, then the left-hand side of
Eq.~\eqref{eq:lb_mom_der} can be identified with the time derivative
$dP/dt=4 \, \hbar \bar{n} \, \dot{\theta}\cos^2\theta$ of the
momentum~\eqref{eq:lb_ds_mom} of the soliton centered around
$\Delta$. Manton's method amounts to identifying, in the right-hand
side of Eq.~\eqref{eq:lb_mom_der}, the contribution due to the soliton
centered at the origin, from which one can infer the force exerted by
one soliton onto the other. Retaining only the leading order of this
contribution and discarding all the other contributions leads to
\begin{equation}
\hbar\, \dot{\theta}\simeq - 8 \, g \bar{n} \cos^4\theta
\exp(-2\Delta\cos\theta/\xi) \, .
\label{eq:theta_der}
\end{equation}
In this formula $\dot{\theta}$ can be expressed in term of the time
derivative of the velocity $V=-c \sin\theta$ of the soliton with
respect to the background. The fact that $V$ changes means that the
soliton under scrutiny does not remain stationary and moves with an
acceleration $\ddot{\Delta}=\dot{V}=-c\,
\dot{\theta}\cos\theta=f(\Delta)$, where $f$ -- which is easily
evaluated from Eq.~\eqref{eq:theta_der} -- is the ratio of the force
experienced by the soliton centered around $\Delta$ to its inertial mass
($m_\mathrm{I}=-4 m \, \xi \bar{n}\, \cos\theta$, see
Ref.~\cite{Pitaevskii2016}). Both the force and the mass are negative
and this results in a repulsive interaction between the solitons.

Once the interaction between two dark solitons has been determined, it is
easy to turn to the case of a chain of solitons, considered as a
one-dimensional lattice of classical particles. For determining the
elementary excitations of such a system, one
writes the spacing between two successive solitons as $\Delta(t) =
\Lambda + X(t)$ and the angular frequency $\Omega$ of the equivalent
spring is just
\begin{equation}
\Omega=\sqrt{- \left.\frac{df}{d\Delta}\right|_{X=0}}
= \frac{4\, c}{\xi} \cos^3\theta\, \exp(-\Lambda\cos\theta/\xi) \, ,
\label{eq:lb_Omega}
\end{equation}
which, together with Eq.~\eqref{eq:lb_freq}, yields the
result~\eqref{eq:fl_low_mode} in the regime where the cnoidal wave
becomes a chain of well-separated solitons. In this regime, the
spacing between nearest solitons being large, the intensity of their
interaction is weak and the lowest branch has a decreasing
amplitude: in the dark-soliton limit of Sec.~\ref{subsec:cn_sol}
it becomes a zero mode corresponding to the translational degree of
freedom of an isolated soliton.

\begin{figure}
\includegraphics[scale=1]{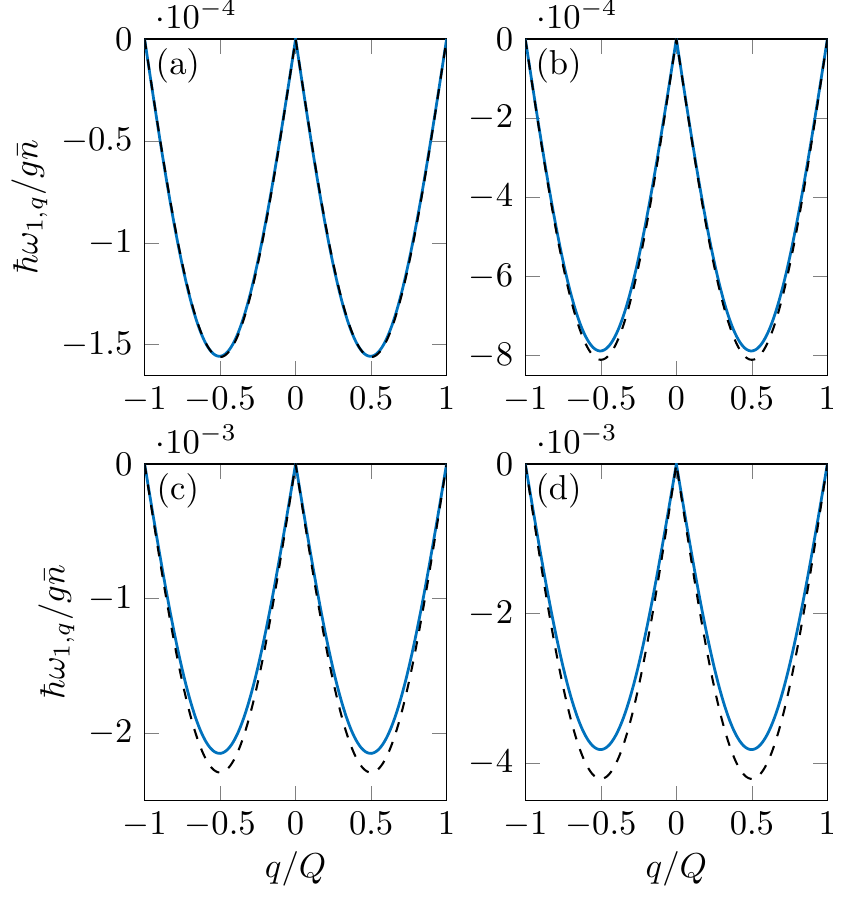}
\caption{Lowest band of the excitation spectrum as a function of $q$.
In each panel we compare the numerical results (blue solid curve) with
the analytic prediction~\eqref{eq:fl_low_mode} (black dashed curve) at
$m_e = 0.99$, i.e., close to the dark-soliton limit. We take $\eta = 0.1$
(a), $0.3$ (b), $0.6$ (c), and $0.9$ (d).}
\label{fig:fl_spectrum_low}
\end{figure}

The accuracy of the approximation~\eqref{eq:fl_low_mode} is
illustrated in Fig.~\ref{fig:fl_spectrum_low} in the case of a cnoidal
wave with $m_e=0.99$, for several values of $\eta$, ranging from $0.1$
to $0.9$. As one can see, the agreement is excellent for low values of
$\eta$ and becomes less accurate when $\eta$ increases. This could be
considered as strange because we repeatedly stated that the validity
of our approximation should only rely on the fact that the separation
$\Lambda$ between two successive solitons is large compared to the
soliton's width $\xi/\cos\theta$, and the ratio of these two
quantities only depends on $m_e$, not on $\eta$ [see
Eq.~\eqref{eq:cn_wavelength}].

\begin{figure}
\includegraphics[scale=1]{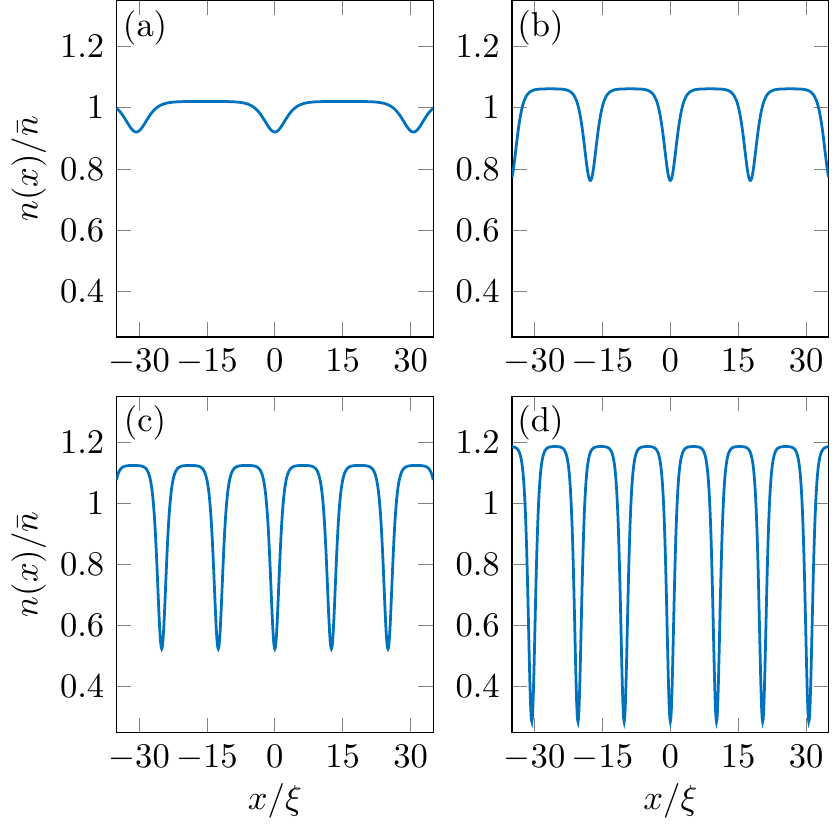}
\caption{Density profiles of the cnoidal wave for $m_e=0.99$ and
different values of $\eta$. The cases (a), (b), (c) and (d)
correspond to the same values of $\eta$ than in
Fig.~\ref{fig:fl_spectrum_low}: $\eta=0.1$, $0.3$, $0.6$ and $0.9$,
respectively.}
\label{fig:fl_dens_prof_low}
\end{figure}

This conundrum is solved by inspecting the two-soliton
Ansatz~\eqref{eq:lb_ansatz}. This Ansatz is valid for evaluating the
interaction between two nearest solitons inasmuch as the ground
state of the train of solitons itself can be described by an
approximate wave function of the type
\begin{equation}
\phi(x)=\sqrt{\bar{n}} \, \prod_{j\in\mathbb{Z}} \Phi(x-j \Lambda) \, .
\label{eq:lb_multi_ds}
\end{equation}
From the density profiles plotted in Fig.~\ref{fig:fl_dens_prof_low},
it is clear that the validity of expression~\eqref{eq:lb_multi_ds} decreases
for increasing values of $\eta$, since the density $n_2$ of the flat
region between two solitons significantly exceeds $\bar{n}$ as $\eta$
increases, contrarily to the situation depicted by Eq.~\eqref{eq:lb_multi_ds}.
From expression~\eqref{eq:cn_roots_2} one sees that, in the regime
$0<1-m_e\ll 1$, $n_2$ remains close to $\bar{n}$ when the additional
condition
\begin{equation}
\eta \ll -\frac{1}{2} \ln(1-m_e)
\label{eq:lb_eta_cond}
\end{equation}
is fulfilled. For the value $m_e=0.99$ corresponding to the plots of
Figs.~\ref{fig:fl_spectrum_low} and~\ref{fig:fl_dens_prof_low}, the
right-hand side of this inequality is equal to $2.3$. This is the reason
why the approximation~\eqref{eq:fl_low_mode} starts being less
accurate when $\eta=0.6$ [see Fig.~\ref{fig:fl_spectrum_low}(c)].

\end{document}